\documentclass[11pt,onecolumn]{IEEEtran}

\usepackage{amsopn}

\newcommand{\Ec}{{\mathcal E}}

\newcommand{\Mc}{{\mathcal M}}

\newcommand{\Tc}{{\mathcal T}}

\newcommand{\Wc}{{\mathcal W}}

\newcommand{\Sv}{{\bf S}}

\newcommand{\Vv}{{\bf V}}
\newcommand{\Wv}{{\bf W}}

\newcommand{\Yv}{{\bf Y}}
\newcommand{\Zv}{{\bf Z}}

\newcommand{\av}{{\bf a}}

\newcommand{\ev}{{\bf e}}

\newcommand{\sv}{{\bf s}}

\newcommand{\onev}{{\bf 1}}
\newcommand{\zerov}{{\bf 0}}

\newcommand{\phiv}{\boldsymbol{\phi}}

\newcommand{\Sh}{{\hat S}}

\newcommand{\Uh}{{\hat U}}

\newcommand{\Xh}{{\hat X}}

\newcommand{\sh}{{\hat s}}

\newcommand{\uh}{{\hat u}}

\newcommand{\xh}{{\hat x}}

\newcommand{\Rt}{{\tilde R}}
\newcommand{\St}{{\tilde S}}

\newcommand{\Zt}{{\tilde Z}}

\def\tr{\mathop{\rm tr}\nolimits}%
\def\Var{\mathop{\rm Var}\nolimits}%
\DeclareMathOperator\E{E}

\newcommand{\Norm}{\mathcal{N}}

\newtheorem{theorem}{Theorem}
\newtheorem{lemma}{Lemma}
\newtheorem{proposition}{Proposition}

\newtheorem{definition}{Definition}
\newtheorem{corollary}{Corollary}

\usepackage{cite}
\usepackage{psfrag}
\usepackage{amsmath}
\usepackage{amsfonts}
\usepackage{graphicx}
\usepackage{version}
\title{Distributed Lossy Averaging}

\author{Han-I~Su,~\IEEEmembership{Student~Member,~IEEE,}
        and~Abbas~El~Gamal,~\IEEEmembership{Fellow,~IEEE}%
\thanks{The authors are with Department of Electrical Engineering,
Stanford University, Stanford, CA 94305, USA.}}

\begin{document}
\maketitle

\begin{abstract}
An information theoretic formulation of the distributed averaging problem previously studied in computer science and control is presented. We assume a network with $m$ nodes each observing a WGN source. The nodes communicate and perform local processing with the goal of computing the average of the sources to within a prescribed mean squared  error distortion. 
The network rate distortion function $R^*(D)$ for a 2-node network with correlated Gaussian sources is established. A general cutset lower bound on $R^*(D)$ is established and shown to be achievable to within a factor of 2 via a centralized protocol over a star network. A lower bound on the network rate  distortion function for distributed weighted-sum protocols, which is larger in order than the cutset bound by a factor of $\log m$ is established. An upper bound on the network rate distortion function for gossip-base weighted-sum protocols, which is only $\log\log m$ larger in order than the lower bound for a complete graph network, is established. The results suggest that using distributed protocols results in a factor of $\log m$ increase in order relative to centralized protocols.  
\end{abstract}


\section{Introduction}

Distributed averaging is a popular example of the distributed consensus problem, which has been receiving much attention recently due to interest in applications ranging from distributed coordination of autonomous agents to distributed computation in sensor networks, ad-hoc networks, and peer-to-peer networks (e.g., see~\cite{JadbabaieLM03,BawaGGM03,GaneshKM03,ScherberP04,BoydGPS05,Olfati-Saber05,DimakisSW06}). 

In this paper, we present a lossy source coding formulation of the distributed averaging problem. We assume that each node in the network observes a source $X_i$, $i=1,2,\ldots,m$, and the nodes communicate and perform local processing with the goal of computing the average $S=(1/m)\sum_{i=1}^mX_i$ to within a prescribed mean squared error distortion. We investigate the network rate distortion function defined as the infimum of average per-node rates that achieve the desired distortion in general and for the class of distributed weighted-sum protocols, which include gossip-based weighted-sum protocols.  Our results, which are information-theoretic, shed light on the fundamental tradeoff in distributed computing between communication and computation accuracy, and the communication penalty of using distributed rather than centralized protocols.  

The paper is organized as follows. In Section~\ref{sec-prelim}, we define the lossy averaging problem and summarize our results. In Section~\ref{sec-review} we briefly review recent work on distributed averaging and discuss how our approach differs from this previous work. 
In Section~\ref{sec-example}, we establish the network rate distortion function for a 2-node network.
In Section~\ref{sec-cutset}, we establish a lower bound on the number of communication rounds needed to achieve a prescribed distortion, and a cutset lower bound on the network rate distortion function. In Section~\ref{sec-gossip}, we investigate the lossy averaging problem for the class of distributed weighted-sum protocols. We establish a lower bound on the network rate distortion function for this class as well as an upper and lower bounds for gossip-based weighted-sum protocols. 

The paper will generally use the notation in~\cite{ElGamalK10}.

\section{Definitions and Summary of Results}
\label{sec-prelim}

Consider a network with $m$ sender-receiver nodes, where node $i=1,2,\ldots,m$ observes an i.i.d. source $X_i$. The nodes communicate and perform local processing with the goal of computing the average of the sources $S=(1/m)\sum_{i=1}^m X_i$ at each node to within a prescribed per-letter distortion $D$. The following definitions apply to any set of correlated i.i.d. sources $(X_1,X_2,\ldots,X_m)$.  In Sections~\ref{sec-cutset} and \ref{sec-gossip}, we assume the sources to be independent white Gaussian noise (WGN) processes each with average power of one.

The topology of the network is specified by a connected graph with no self loops $(\Mc,\Ec)$, where $\Mc=\{1,2,\ldots,m\}$ is the set of nodes and $\Ec$ is a set of undirected edges (node pairs) $\{i,j\}$, $i,j \in \Mc$ and $i\ne j$. Communication is performed in {\em rounds} and each round is divided into {\em time slots}. Each round may consist of a different number of time slots, and each time slot may consist of a different number of transmissions. One edge (node pair) is chosen in each round and only one node is allowed to transmit in each time slot. Although in general multiple edges may be chosen in the same round and multiple nodes may be allowed to transmit in a time slot, we restrict ourselves to one edge and one node at a time to simplify the analysis. Without loss of generality we assume that the selected node pair communicate in a round robin manner with the first node communicating in odd time slots and the second node communicating in even time slots. Further, 
 we assume a {\em source coding} setting, where communication is noiseless and instant, that is, every transmitted message is successfully received by the intended receiver in the same time slot it is transmitted in. As in most multiple-user information theoretic setups, we assume that each node has a sequence of $n$ symbols of its source before communication and computing commences. We seek to find the limit on the tradeoff between communication and distortion as $n$ tends to infinity.

Communication and local computing in the network are performed according to an agreed upon $(T,p(\ev),R,n)$ {\em averaging protocol} that consists of:
\begin{enumerate}
\item The number of communication rounds $T$. 
\item A probability mass function $p(\ev)$ for $\ev \in\Ec_1^T$, where $\Ec_1^T=\{(e_1,e_2,\ldots,e_T): e_t\in\Ec, t=1,2,\ldots,T\}$ is the set of all feasible edge selection sequences. 
\item A set of $(\ev,R(\ev),n)$ block codes, one for each edge selection sequence $\ev\in\Ec_1^T$. Each $(\ev,R(\ev),n)$ block code consists of:
\begin{enumerate}
\item A set of encoding functions, one for each node in each round and each time slot. Suppose that in round $t\in[1:T] =\{1,2,\ldots,T\}$, edge $e_t=\{i,j\}$ is selected, and this round consists of $q_t\ge 1$ time slots. Without loss of generality, assume that $q_t$ is even, node $i$ transmits in odd time slots, and node $j$ transmits in even time slots. In time slot $\nu \in\{1,3,\ldots,q_t-1\}$, node $i$ sends an index $w_{t\nu}(x_{i1}^n,w_{t1}^{\nu-1},\Wc_{i1}^{t-1}) \in [1:2^{n r_{t\nu}}]$, where $\Wc_{i\tau} = \{w_{\tau1}^{q_\tau}: i\in e_\tau\}$, and thus $\Wc_{i1}^{t-1}$ is the collection of indices node $i$ has up to time $t-1$. Similarly, node $j$ sends an index $w_{t\nu}(x_{j1}^n,w_{t1}^{\nu-1},\Wc_{j1}^{t-1}) \in [1:2^{nr_{t\nu}}]$ in time slot $\nu\in\{2,4,\ldots,q_t\}$. The total transmission rate per source symbol of node $i$ in round $t$ is
\[
r_i(t) = \sum_{\nu \in \{1,3,\ldots,q_t-1\}} r_{t\nu},
\] 
and similarly for node $j$, the total transmission rate is 
$r_j(t) = \sum_{\nu\in\{2,4,\ldots,q_t\}} r_{t\nu}$. 

\item A set of decoding functions, one for each node. At the end of round $T$, the decoder for node $i\in\Mc$ assigns an estimate $s_{i1}^n(x_{i1}^n,\Wc_{i1}^T) = (s_{i1}(x_{i1}^n,\Wc_{i1}^T), s_{i2}(x_{i1}^n,\Wc_{i1}^T), \ldots, s_{in}(x_{i1}^n,\Wc_{i1}^T))$ of the average $s_1^n =(s_1,s_2,\ldots,s_n)$ to each source sequence and all messages received by the node, where $s_k=(1/m)\sum_{i=1}^m x_{ik}$. 
\end{enumerate}
\end{enumerate}
Let $r_i(t)=0$ if node $i$ is not selected in round $t$. Then the total transmission rate for node $i\in\Mc$ is defined as
\[
R_i = \sum_{t=1}^T r_i(t).
\]
The per-node transmission rate is defined as
\[
R(\ev) = \sum_{i=1}^m \frac{1}{m}R_i,
\]
and the average per-letter distortion is defined as
\[
D(\ev) = \frac{1}{mn} \sum_{i=1}^m \sum_{k=1}^n \E\left( (S_k-S_{ik})^2 \right),
\]
where $S_k=(1/m)\sum_{i=1}^m X_{ik}$ and the expectation is taken over the source statistics. Finally, the expected per-node transmission rate is defined as
\[
R = \sum_{\ev\in\Ec_1^T} p(\ev) R(\ev).
\]
For a fixed $T$ and $p(\ev)$, a rate distortion pair $(R,D)$ is said to be {\em achievable} if there exists a sequence of $(T,p(\ev),R,n)$ code sets indexed by the block length $n$ such that
\[
\limsup_{n \to \infty} \sum_{\ev\in\Ec_1^T} p(\ev) D(\ev) \le D.
\]
The {\em network rate distortion function} $R^*(D)$ is defined as the infimum over the number of rounds $T$ and probability mass functions $p(\ev)$ of all rates $R$ such that the pairs $(R,D)$ are achievable. 

\medskip
\noindent{\em Centralized versus Distributed Protocols}: A goal of our work is to quantify the communication penalty of using distributed relative to centralized protocols. In a distributed protocol, such as the distributed weighted-sum protocol discussed in Section~\ref{sec-gossip}, the code used at each round is the same for all nodes, that is, it does not depend on the identities of the selected nodes. The code, however, can be time-varying, that is, can change with the round number. In a centralized protocol, on the other hand, the code can depend on the node identities. For example, a node may be designated as a ``cluster-head'' and treated differently by the centralized protocol than other nodes.
\medskip

As an example of distributed average protocols, we consider a class of {\em distributed weighted-sum protocols}. Before defining this class of protocols, we briefly review rate distortion theory for a WGN source and the mean squared error distortion measure~\cite{CoverT06}. In this setup, sender node 1 has a WGN source $X$ with average power $P$ and wishes to send a description $\Xh$ of the source to node 2 with normalized distortion $d=D/P\in [0,1]$. We assume standard definitions for a code, distortion, achievability, and rate distortion function. Then, the  rate distortion function is
\[
R(D) = \min_{P(\xh|x): \E\left( (X-\Xh)^2 \right)\le Pd} I(X;\Xh) = \frac{1}{2}\log\frac{1}{d}.
\]
The test channel $P(\xh|x)$ that achieves the minimum can be expressed as
\begin{align}
\label{equ-test-ch}
\Xh = (1-d)(X+Z),
\end{align}
where $Z\sim\Norm(0,Pd/(1-d))$ is independent of $X$. The  rate distortion function is achieved by using a sequence of codebooks $\{\xh^n(w):\; w \in [1:2^{nR}]\}$, where each estimate (description) $\xh^n(w)$ is generated independently according to an i.i.d. $\Norm(0,P(1-d))$ distribution, and joint typicality encoding~\cite{CoverT06}. We refer to such codes as \emph{Gaussian codes}.
  
We are now ready to define the class of distributed weighted-sum protocols. We assume that the sources $X_1,X_2,\ldots,X_m$ are independent WGN processes, each with average power one.  A {\em distributed weighted-sum protocol} is characterized by $(T,p(\ev),R,n,d)$, where $T$, $p(\ev)$, $R$, and $n$ are defined as before, and $d\in [0,1]$ is a \emph{normalized local distortion}. Let $S_i(0)=X_i$ for $i \in \Mc$ and fix $T$, $d$, and an edge selection sequence $\ev \in \Ec_1^T$. Assuming edge $\{i,j\}$ is selected in round $(t+1)$, define the test channels
\begin{align*}
\Sh_i(t) &= (1-d)(S_i(t)+Z_i(t)),\\
\Sh_j(t) &= (1-d)(S_j(t)+Z_j(t)),
\end{align*}
where $Z_i(t)$ and $Z_j(t)$ are independent WGN sources with average powers $\E(S_i(t)^2)d/(1-d)$ and $\E(S_j(t)^2)d/(1-d)$, respectively, and they are independent of $(X_1,X_2,\ldots,X_m)$ and $\{Z_l(\tau): l\in e_\tau, \tau\in[0:t-1]\}$. Then the expected distortion between $S_i(t)$ and the output of the test channel $\Sh_i(t)$ is 
\[
\E\left( (S_i(t)-\Sh_i(t))^2 \right) = \E(S_i(t)^2) d.
\]
Similarly, the expected distortion between $S_j(t)$ and $\Sh_j(t)$ is $\E(S_j(t)^2)d$. Now, define the updated sources
\begin{equation}
\label{equ-gossip-s}
\begin{array}{c}
\displaystyle S_{i}(t+1) = \frac{1}{2} S_{i}(t) + \frac{1}{2(1-d)}\Sh_{j}(t), \\[8pt]
\displaystyle S_{j}(t+1) = \frac{1}{2} S_{j}(t) + \frac{1}{2(1-d)}\Sh_{i}(t),
\end{array}
\end{equation}
and $S_l(t+1) = S_l(t)$ for $l \in \Mc\setminus \{i,j\}$. Note that since $\{S_i(0): i \in \Mc\}$ and $\{Z_l(t): l \in e_t, t\in [0:T-1]\}$ are independent Gaussian and the test channels and update equations are linear, $S_i(t)$ is Gaussian for every $i \in \Mc$ and $t\in [1:T]$. These Gaussian test channels are then used to generate Gaussian codes that are revealed to all parties prior to network operation. 

Now we describe the $(\ev,R(\ev),n)$ block codes for a distributed weighted-sum protocol. Initially, each node $i\in \Mc$ has an estimate $s_{i1}^n(0) =x_{i1}^n$ of the true average $s_1^n$. In each round, communication is performed independently in two time slots. Assume that edge $\{i,j\}$ is selected in round $t+1$. In the first time slot node $i$ uses its Gaussian codes to describe the source $S_i(t)$ to node $j$. In the second time slot, node $j$ similarly describes $S_j(t)$ to node $i$ using its Gaussian codes. At the end of the second time slot, nodes $i$ and $j$ compute the updated estimates
\begin{equation}
\label{equ-gossip-si}
\begin{array}{c}
\displaystyle s_{i1}^n(t+1) = \frac{1}{2} s_{i1}^n(t) + \frac{1}{2(1-d)}\sh_{j1}^n(t), \\[8pt]
\displaystyle s_{j1}^n(t+1) = \frac{1}{2} s_{j1}^n(t) + \frac{1}{2(1-d)}\sh_{i1}^n(t),
\end{array}
\end{equation}
respectively, where $\sh_{i1}^n(t)$ and $\sh_{j1}^n(t)$ are the estimates (descriptions) of $s_{i1}^n(t)$ and $s_{j1}^n(t)$. The estimate for node $l \in \Mc\setminus\{i,j\}$ remains unchanged, that is, $s_{l1}^n(t+1)=s_{l1}^n(t)$.

At the end of round $T$, node $i \in \Mc$ sets its final estimate of the average as $s_{i1}^n= s_{i1}^n(T)$ if it is involved in at least one round of communication, otherwise it sets $s_{i1}^n=(1/m)s_{i1}^n(0)$. Thus for the degenerate distributed weighted-sum protocol with $T=0$, node $i \in \Mc$ sets its final estimate to $(1/m)s_{i1}^n(0)$. 

We define the {\em weighted-sum network rate distortion function} $R^*_{\rm WS}(D)$ in the same manner as $R^*(D)$ except that the codes used are restricted to the above class.

\medskip
\noindent{\em Remarks:} 
\begin{enumerate}
\item The weights in the update equations~\eqref{equ-gossip-s} are chosen such that source $S_i(t)$ is a sum of a convex combination of $(X_1,X_2,\ldots,X_m)$ with coefficients independent of the normalized local distortion $d$ and an independent noise. Note that as we let $d$ approach to zero, the update equations for the distributed weighted-sum protocol reduce to those for the standard gossip algorithm~\cite{BoydGPS05}.

\item The distributed weighted-sum protocol as defined above does not exploit the build up in correlation between the node estimates induced by communication and local computing. This correlation can be readily used to reduce the rate via Wyner-Ziv coding. However, we are not able to obtain general upper and lower bounds on the network rate distortion function with side information because the correlations between the estimates are time varying and depend on the particular edge selection sequence. We explore the rate reduction achieved by leveraging this correlation through simulations.
\end{enumerate}
\medskip

We also consider {\em gossip-based weighted-sum protocols} $(T,Q,R,n,d)$, where $Q$ is an $m\times m$ stochastic matrix such that $Q_{ij}=0$ if $\{i,j\}\notin \Ec$. In each round of a gossip-based weighted-sum protocol, a node $i$ is selected uniformly at random from $\Mc$. Node $i$ then selects a neighbor $j\in \{j: \{i,j\}\in \Ec\}$ with conditional probability $Q_{ij}$. Note that this edge selection process is the same as the asynchronous time model in~\cite{BoydGPS05}. After the edge (node pair) $\{i,j\}$ is selected, the code for the weighted-sum protocols described above is used. Thus, a gossip-based weighted-sum protocol $(T,Q,R,n,d)$ is a distributed weighted sum protocol $(T,p(\ev),R,n,d)$ with 
\[
p(\ev) = \frac{1}{m^T} \prod_{\{i,j\}\in\Ec} Q_{ij}^{|\{t\in[1:T]: e_t=\{i,j\}\}|}
\]
for every $\ev\in\Ec_1^T$. We define the {\em gossip-based network rate distortion function} $R^*_\mathrm{GWS}(D)$ for the class of gossip-based weighted-sum protocols in a similar way as $R^*_\mathrm{WS}(D)$. Note that $R^*(D) \le R^*_\mathrm{WS}(D) \le R^*_\mathrm{GWS}(D)$.

The following result will be used in the bounds on $R^*(D)$, $R^*_\mathrm{WS}$, and $R^*_\mathrm{GWS}(D)$.

\medskip
\begin{lemma}
\label{lm-achv}
Consider the distributed weighted-sum protocol defined above for some fixed $T$, $d$, and edge selection sequence $\ev \in \Ec_1^T$. Then the distortion $\E((S_i(t))^2) d$ between $S_i(t)$ and $\Sh_i(t)$ is achievable if
\begin{align*}
r_i(t+1) &\ge \frac{1}{2}\log\frac{1}{d}
\end{align*}
for every $i \in e_{t+1}$ and $t \in [0:T-1]$.
\end{lemma}
\medskip

\begin{proof}
The lemma can be proved by the procedure in~\cite{ElGamalK10} for extending achievability of a lossy source coding problem for finite sources and distortion measures to Gaussian sources with mean squared error distortion. We first establish achievability for finitely quantized versions $[S_i(t)]$ of $S_i(t)$ for $i \in \Mc$ and $t\in[0:T]$ and $[\Sh_i(t)]$ of $\Sh_i(t)$ for $i \in \Mc$ and $t\in[0:T-1]$. We then use the covering lemma, the joint typicality lemma, and the Markov lemma, to show that joint typicality encoding succeeds with high probability if
\begin{align*}
r_i(t+1) &\ge I([S_i(t)];[\Sh_i(t)])
\end{align*}
for every $i \in e_{t+1}$ and $t \in [0:T-1]$. When encoding succeeds, the  distortion between $S_{i1}^n(t)$ and its description $\Sh_{i1}^n(t)$ is close to $\E([S_i(t)]^2) d$ for $i \in e_{t+1}$ and $t \in [0:T-1]$. Finally, taking appropriate limits, it can be readily shown that the distortion $\E((S_i(t))^2) d$ is achievable for the Gaussian sources and descriptions if
\begin{align*}
r_i(t+1) &\ge \frac{1}{2}\log\frac{1}{d},
\end{align*}
for every $i \in e_{t+1}$ and $t \in [0:T-1]$.
\end{proof}

\subsection{Summary of Results}

\noindent{\em Section~\ref{sec-example}}: We establish $R^*(D)$ for a 2-node network with correlated Gaussian sources (Proposition~\ref{lm-2node-lower}).
\medskip

\noindent{\em Section~\ref{sec-cutset}}: 
\begin{enumerate}
\item We establish a lower bound on the number of rounds needed to achieve distortion $D< 1/m^3$, $T\ge 2m-3$ (Proposition~\ref{prop-min-round}).
\item For independent WGN sources, we establish the following cutset lower bound on the network rate distortion function (Theorem~\ref{thm-cutset})
\[
R^*(D)\ge \frac{1}{2} \log \frac{(m-1)}{m^2D}.
\]
The bound is tight for $m=2$.
\item We show that a centralized protocol over a star network can achieve rates within a factor of 2 of the cutset bound for large $m$ (Proposition~\ref{prop-star}). We establish a tighter cutset bound for this network and show that it becomes tight as $D \to 0$.  
\end{enumerate}
\medskip

\noindent{\em Section~\ref{sec-gossip}}: We investigate a  class of distributed weighted-sum protocols, including gossip-based weighted-sum protocols for independent WGN sources. 
\begin{enumerate}
\item We establish the following lower bound on the network rate distortion function for distributed weighted-sum protocols (Theorem~\ref{thm-dist-lower})
\[
R^*_\mathrm{WS}(D)\ge \frac{1}{2} \left(\log \frac{1}{\sqrt{D}+1/m}\right) \left(\log \frac{1}{4mD}\right).
\]
This bound is larger than the cutset bound by a factor of $\log m$ in order.

\item We establish the following bounds on the expected network rate distortion function for a class of gossip-based weighted-sum protocols (Theorem~\ref{thm-dist-upper})
\begin{align*}
R^*_\mathrm{GWS}(D) &= \Omega\left( \left(\log\frac{1}{D}\right)\left(\log\frac{1}{mD}\right)\right), \\
R^*_\mathrm{GWS}(D) &= O\left( 
\frac{1}{m(1-\lambda_2)}\left(\log\frac{1}{D}\right)\left(\log \log \frac{1}{D} + \log\frac{1}{m^2 (1-\lambda_2)D}\right)\right),
\end{align*}
where $\lambda_2$ is the second largest eigenvalue of the expected averaging matrix~\cite{BoydGPS05}. The upper bound is shown to hold for general independent sources, i.e., not necessarily Gaussian. For distortion $D=o(1/m\log m)$, the upper bound has the same order as the lower bound, while for distortion $D=\Omega(1/m\log m)$, it is larger in order than the lower bound by a factor of $\log\log m$.  Since a centralized protocol can achieve the same order as the cutset bound, and the lower bound on distributed weighted-sum protocols is $\log m$ larger in order than the cutset bound and is achievable to within order $\log\log m$, the order $\log m$ factor represents the penalty of using distributed protocols relative to centralized protocols.
\item We use simulations to explore the improvement in rate achieved by exploiting the correlation induced by communication and local computing. 
\end{enumerate}

\section{Relationship to Previous Work}
\label{sec-review}
Examples of work on distributed averaging under the synchronous model include~\cite{XiaoB03,OlshevskyT06}, where deterministic linear iterative protocols are used. Each node iteratively computes the weighted sum of the estimates of its neighbors and itself, that is, $\sv(t+1) = A\sv(t)$, where $A$ is a nonnegative matrix with nonzero entries $a_{ij}$ only if there is an edge between nodes $i$ and $j$. The results in \cite{XiaoB03} show that when $A$ is a doubly stochastic matrix, the network achieves consensus average as $t\to\infty$. Furthermore, when the topology of entire network is known, the optimal $A$ that achieves the fastest convergence can be computed via semidefinite programming. In \cite{OlshevskyT06}, it is shown that if $A$ is a stochastic matrix, the states converge to a common weighted-sum of initial states. The weights correspond to the steady-state probabilities of the Markov chain associated with the stochastic matrix. If the initial states are divided by the corresponding weights, the consensus average can be reached.

Synchronous protocols cannot be used in networks with link failures or dynamic topologies. This has motivated the development of the gossip protocol, which was first introduced for computing the average at a sink node in a peer-to-peer network~\cite{BawaGGM03} and later applied to distributed averaging (e.g.,~\cite{ BoydGPS05}). 
In each communication round, a node and one of its neighbors are selected. The two nodes update their estimates by averaging their current estimates. Note that this process does not change the average of the states in the network. In \cite{KempeDG03}, it is assumed that in each communication round, a node and its neighbor are selected uniformly at random. The results provide order bounds on convergence time that hold with high probability. In~\cite{BoydGPS05}, nodes select neighbors to communicate with according to a doubly stochastic matrix. 
Bounds on convergence time that hold with high probability are obtained as a function of the second largest eigenvalue of the matrix. The node selection statistics that minimize the second largest eigenvalue can be found by a distributed subgradient method. Motivated by wireless networks and the Internet, the paper also investigates distributed averaging for networks modeled by geometric random graphs and by preferential connectivity. It is shown that convergence time under the preferential connectivity does not depend on the size of the network.

In \cite{OlshevskyT06}, a variation of the gossip-type protocol is studied for a network with link failures and dynamic topologies. In each communication round, each node first broadcasts its current estimate to its neighbors. Node $i$ then makes an offer to neighbor $j$ if $s_j(t) < s_i(t)$ and $s_j(t)$ is smaller than the states of other neighbors of node $i$. At the end of this round, each node accepts the offer from the node with largest estimate, and both nodes update their estimates with their averages. It is proved in~\cite{OlshevskyT06} that this protocol converges under certain connectivity constraints.

The aforementioned work involves the noiseless communication and computation of real numbers, which is unrealistic. The effect of quantization on distributed consensus has received attention only recently. In \cite{XiaoBK07}, quantization noise is modeled as additive white noise. It is shown that the expectation of the state vector converges to the average of the initial states, but the variance diverges. Further, it is shown that the mean square deviation of the state vector is bounded away from zero. 
The tradeoff between mean square deviation and convergence time is also investigated. 
Recognizing that the divergence of the consensus variance in \cite{XiaoBK07} is due to the assumption that quantization noise is white, the work in~\cite{YildizS08} exploits the time and spatial correlation of the estimates of the nodes. The initial states are assumed to be random variables with zero mean and finite variance. A differential nested lattice encoding quantizer that combines predictive coding and Wyner-Ziv coding is used. At each round, node $i$ updates its estimate with a weighted-sum of the estimates of its neighbors and itself and an additive quantization noise, hence the estimates $s_i(t)$ and $s_i(t+1)$ are correlated. The time correlation is exploited by predictive coding to reduce quantization error. The update process also increases the spatial correlation between nodes. As such, the estimate of each node is used as side information to reconstruct descriptions from received quantized indices. It is shown that the mean squared error is bounded when the optimal lattice vector quantizer is used, and the transmission rate at round $t$ approaches zero as $t\to\infty$. The tradeoff between the rate per node per round and the final mean squared error is also investigated through simulations.

The work in~\cite{KashyapBS07} and \cite{AysalCR07} use a different approach to quantized consensus. Each node has an integer-valued initial estimate and the nodes wish to compute the quantized average consensus, which is reached when $s_i(t) \in \{\lfloor s \rfloor, \lceil s \rceil\}$ for $i = 1,2,\ldots,m$, where $s$ is the average of the initial estimates. It is shown that simply uniformly quantizing the estimates of the gossip-based protocol is not sufficient to achieve quantized average consensus and a gossip-based protocol is shown to achieve it. In \cite{AysalCR07} the integer-valued averaging problem with exact consensus is investigated. A probabilistic quantizer is added to the real-valued gossip-based protocol. The estimate $s_i(t)$ is quantized either to $u = \lceil s_i(t) \rceil$ or $l = \lfloor s_i(t) \rfloor$ with probabilities $(u - s_i(t))$ and $(s_i(t) - l)$, respectively. The results show that the expectation of the common estimate is the average $s$. In~\cite{CarliFFZ10}, different update rules for achieving quantized consensus by using deterministic uniform quantizers and probabilistic quantizers are compared. In~\cite{CarliZ07,NedicOOT07}, exchanging and storing quantized information is considered for the consensus problem.  

The first information theoretic work on distributed averaging is reported in~\cite{AyasoSD08}. The nodes communicate through channels with finite capacities. Each node is required to compute a function of the initial values to within a desired mean squared error distortion. Lower and upper bounds on the time to achieve the desired distortion are shown to be inversely proportional to the graph conductance.

Our work is related most closely to the work on quantized averaging in~\cite{XiaoBK07,YildizS08} and the information theoretic work in~\cite{AyasoSD08}. Compared to previous work on quantized averaging, our information-theoretic approach to the problem deals naturally and fundamentally with quantization and the results provide limits that hold independent of implementation details. Our results are difficult to compare with the results in these papers, however, because of basic differences in the models and assumptions. While the work in~\cite{AyasoSD08} is information theoretic, it deals with a different formulation than ours and the results are not comparable. Our formulation of the distributed averaging problem can be viewed also as a generalization of the CEO problem~\cite{PrabhakaranTR04,Oohama05},  where in our setting every node wants to compute the average and the communication protocol is significantly more complex in that it allows for interactivity, relaying, and local computing, in addition to multiple access. 

\section{$R^*(D)$ for 2-Node Network}
\label{sec-example}

Consider a network with 2 nodes and a single edge, and assume correlated WGN sources $(X_1,X_2)$ with covariance matrix
\[
K = \left[ \begin{array}{cc}
P_1 & \rho\sqrt{P_1P_2} \\ \rho\sqrt{P_1P_2} & P_2
\end{array} \right].
\]
Each node wishes to compute the weighted sum $g(X_1,X_2) = a_1X_1+a_2X_2$, for some constants $a_1$ and $a_2$, to within mean squared error distortion $D$.

For a 2-node network, there is only one round of communication with an arbitrary number of time slots. The interesting case is when distortion is small enough such that each node must transmit to the other node. The following proposition establishes the network rate distortion function for this regime.

\medskip
\begin{proposition} 
\label{lm-2node-lower}
The network rate distortion function for the 2-node network with correlated WGN sources is
\[
R^*(D) = \frac{1}{2}\log\left( \frac{a_1a_2(1-\rho^2)\sqrt{P_1P_2}}{D}\right)
\]
for $D<\min\left\{a_1^2(1-\rho^2)P_1,a_2^2(1-\rho^2)P_2\right\}$.
\end{proposition}
\medskip

\begin{IEEEproof}
The converse follows by a cutset bound argument given in the Appendix.

Achievability of the network rate distortion function follows by performing two independent Wyner-Ziv coding~\cite{WynerZ76} steps. In the first time slot, node 1 uses Wyner-Ziv coding to describe its source $X_1$ to node 2 at rate $R_1 = (1/2) \log(a_1^2(1-\rho^2)P_1/D)$. In the second time slot, node 2 uses the Wyner-Ziv coding to describe its source $X_2$ to node 1 at rate $R_2= (1/2) \log (a_2^2(1-\rho^2)P_2/D)$. At the end of the second time slot, nodes 1 and 2 compute the estimates 
\[
g_{11}^n = a_1x_{11}^n + a_2\xh_{21}^n\ \text{and}\ g_{21}^n = a_1\xh_{11}^n + a_2x_{21}^n,
\]
where $\xh_{11}^n$ and $\xh_{21}^n$ are the descriptions of $x_{11}^n$ and $x_{21}^n$, respectively. The average per-letter distortion between the estimates and the objective weighted-sum is 
\begin{align*}
&\lim_{n\to\infty} \frac{1}{2n}\sum_{i=1}^2\sum_{k=1}^n\E\left((g(X_{1k},X_{2k}) - G_{ik})^2\right) \\
&\qquad= \lim_{n\to\infty} \frac{1}{2n}\sum_{k=1}^n\left(a_2^2\E\left((X_{2k} - \Xh_{2k})^2\right) + a_1^2\E\left((X_{1k}-\Xh_{1k})^2\right)\right) \\
&\qquad= \frac{1}{2}\left( a_2^2\frac{D}{a_2^2} + a_1^2\frac{D}{a_1^2} \right)= D.
\end{align*}
The per-node transmission rate is
\[
R = \frac{R_1+ R_2}{2} = \frac{1}{2}\left(\frac{1}{2}\log\frac{a_1^2(1-\rho^2)P_1}{D}+\frac{1}{2}\log\frac{a_2^2(1-\rho^2)P_2}{D}\right) 
= \frac{1}{2}\log\left( \frac{a_1a_2(1-\rho^2)\sqrt{P_1P_2}}{D}\right).
\]
This completes the proof of the proposition.
\end{IEEEproof}
\medskip

\noindent{\em Remarks}:
\begin{enumerate}
\item In~\cite{Kaspi85}, Kaspi investigated the interactive lossy source coding problem when the objective is for each source to obtain a description of the other source. Our problem is different and as such Kaspi's results do not readily apply. In~\cite{MaI08}, the interactive communication problem for asymptotically lossless computation is investigated. Again their results do not apply to our setting because they do not consider loss.
\item In the above example, it is optimal for each node to simply compress its own source and send the compressed version to the other node, that is, only two time slots and no intermediate computing are necessary and sufficient. Based on the results in ~\cite{Kaspi85,MaI08}, we do not expect these conclusions to hold in general for non-Gaussian sources, other distortion measures, and other functions.
\item Finding the rate distortion function for a 3-node network even with Gaussian sources is difficult because (i) there can be several possible feasible edge selection sequences and it is not a priori clear which sequence yields the optimal per-node rate, (ii) the codes allow relaying in addition to interactive communication and local computing, and (iii) it is not known if Gaussian codes are optimal. Results on a 3-node problem are reported in~\cite{CuffSE09}.  
\end{enumerate}

\section{Lower Bound on $T$ and $R^*(D)$}
\label{sec-cutset}

In this section, we establish a general lower bound on the minimum number of rounds $T$ for any averaging protocol. We then establish a cutset bound on $R^*(D)$ for independent WGN sources each with average power one. Finally, we show that the cutset bound can be achieved within a factor of 2 via a centralized protocol over a star network. 

\subsection{Lower bound on $T$}
We first establish the following lower bound on the minimum number of rounds needed by any averaging protocol.

\medskip
\begin{proposition}
\label{prop-min-round}
Every averaging protocol that achieves distortion $D< 1/m^3$ must use at least $T \ge 2m-3$ rounds. 
\end{proposition}
\medskip

\begin{IEEEproof}
Let $\Ec_t$ be the set of edges selected in rounds $1,2,\ldots,t$. Suppose that the graph $(\Mc,\Ec_t)$ is not connected. Then, for each node $i\in \Mc$, there exists a node $j(i)$ such that there is no path between these two nodes. The estimate $S_{i1}^n$ of node $i$ is independent of the source $X_{j(i),1}^n$ and its distortion is then lower bounded by
\[
\frac{1}{n} \sum_{k=1}^n \E\left((S_k-S_{ik})^2\right) \ge \frac{1}{n} \sum_{k=1}^n \E\left(\left(\frac{X_{j(i),k}}{m}\right)^2\right) = \frac{1}{m^2} > D.
\]
Thus, if distortion $D< 1/m^3$ is to be achieved at the end of round $t$, the graph $(\Mc,\Ec_t)$  must be connected and $t\ge m-1$.

Now let $t\ge m-1$ be the smallest round index such that the graph $(\Mc,\Ec_t)$ is connected. If the number of rounds $T < t+m-2$, then there exists at least one node $i \in \Mc\setminus e_t$, which is not selected after round $t-1$. However, the graph $(\Mc,\Ec_{t-1})$ is not connected,  and thus the estimate of node $i$ must have distortion higher than $1/m^2$. Then the distortion $D < 1/m^3$ cannot be achieved. Therefore, the number of rounds to achieve distortion $D<1/m^3$ is lower bounded by
\[
T \ge t+m-2 \ge 2m-3.
\]
\end{IEEEproof}

\medskip
\noindent{\em Remark:}  The averaging protocol that achieves $R^*(D)$ does not necessarily have to use the smallest $T$. It may be possible to use less bits  by using more rounds. We do not, however, have a specific example of such case.

\subsection{Cutset Bound}

Consider the $m$-node distributed lossy averaging problem when the sources $(X_1,X_2,\ldots,X_m)$ are independent WGN processes each with average power one. We establish the following cutset lower bound on the network rate distortion function. 

\begin{theorem}
\label{thm-cutset}
The network rate distortion function $R^*(D)=0$ if $D \ge (m-1)/m^2$ and is lower bounded by
\begin{equation*}
R^*(D) \ge \frac{1}{2}\log\left(\frac{m-1}{m^2D}\right) \text{ if } D< \frac{m-1}{m^2}.
\end{equation*} 

\end{theorem}

\begin{IEEEproof}
With no communication, i.e., $R=0$, the best estimate of each node is the MMSE estimate 
$S_{ik} = \E(S_k|X_{i1}^n) = X_{ik}/m$ for $k \in [1:n]$ and $i \in \Mc$. Let $U_{ik} = (1/m)\sum_{j \in \Mc\setminus\{i\}} X_{jk}$. Then the distortion in this case is  
\[
\frac{1}{mn}\sum_{i=1}^m\sum_{k=1}^n\E\left((S_k-S_{ik})^2\right) = \frac{1}{mn}\sum_{i=1}^m\sum_{k=1}^n \E\left(U_{ik}^2\right) = \frac{m-1}{m^2}.
\]
Thus, if $D\ge (m-1)/m^2$, $R^*(D) = 0$.

Next we consider $D < (m-1)/m^2$. Fix $T$ and $\ev\in\Ec_1^T$ and consider an $(\ev,R(\ev),n)$ block code that achieves distortion $D(\ev)$. Since only pairwise communications are allowed, the number of bits transmitted by all nodes is equal to the number of bits received by all nodes. We consider the number of bits received by node $i$, denoted by $n\Rt_i(\ev)$. Let $W_i$ be the collection of indices sent from nodes $j \in \Mc\setminus\{i\}$ to node $i$. Then the estimate $S_{i1}^n$ of node $i$ is a function of its source $X_{i1}^n$ and the received message $W_i$. We can bound the receiving rate as follows 
\begin{align*}
n\Rt_i(\ev) &\ge H(W_i) \ge H(W_i|X_{i1}^n) \ge I(U_{i1}^n;W_i|X_{i1}^n) \\
&= \sum_{k=1}^n \left( h(U_{ik}|U_{i1}^{k-1},X_{i1}^n)-h(U_{ik}|U_{i1}^{k-1},X_{i1}^n,W_i) \right) \\
&= \sum_{k=1}^n \left( h(U_{ik})-h(U_{ik}|U_{i1}^{k-1},X_{i1}^n,W_i,S_{i1}^n) \right) \\
&\ge \sum_{k=1}^n \left( h(U_{ik})-h(U_{ik}|X_{ik},S_{ik}) \right) \\
&= \frac{n}{2}\log\left(2\pi e\frac{m-1}{m^2}\right) - \sum_{k=1}^n h\left(\left. U_{ik} + \frac{1}{m}X_{ik}-S_{ik}\right|X_{ik},S_{ik}\right) \\
&\ge \frac{n}{2}\log\left(2\pi e\frac{m-1}{m^2}\right) - \sum_{k=1}^n h(S_k-S_{ik}) \\
&\ge \frac{n}{2}\log\left(\frac{m-1}{m^2D_i}\right),
\end{align*}
where $D_i = (1/n)\sum_{k=1}^n\E\left((S_{k}-S_{ik})^2\right)$. The per-node transmission rate is lower bounded by
\[
R(\ev) \ge \min_{(1/m)\sum_{i=1}^mD_i \le D(\ev)} \left(\frac{1}{m}\sum_{i=1}^m \frac{1}{2}\log\left(\frac{m-1}{m^2D_i}\right)\right) 
= \frac{1}{2}\log\left(\frac{m-1}{m^2D(\ev)}\right),
\]
which follows from Jensen's inequality and the distortion constraint $(1/m)\sum_{i=1}^m D_i \le D(\ev)$. For any probability mass function $p(\ev)$ such that $\sum_{\ev \in \Ec_1^T} p(\ev)D(\ev) \le D$, the expected per-node transmission rate is lower bounded by
\[
\sum_{\ev \in \Ec_1^T} p(\ev) R(\ev) \ge \sum_{\ev \in \Ec_1^T} p(\ev) \left(  \frac{1}{2}\log\frac{m-1}{m^2D(\ev)}\right) 
\ge \frac{1}{2}\log\left(\frac{m-1}{m^2D}\right).
\]
Since $T$ is arbitrary, the network rate distortion function is also lower bounded by
\[
R^*(D) \ge \frac{1}{2}\log\left(\frac{m-1}{m^2D}\right)\ \text{for}\ D<\frac{m-1}{m^2}.
\]
\end{IEEEproof}
\medskip

\noindent{\em Remarks}: 
\begin{enumerate}
\item  As can be be readily verified from Proposition~\ref{lm-2node-lower}, the above lower bound is achieved for $m=2$.  
\item In the following subsection, we show that a centralized protocol for a ``star'' network can achieve a rate less than twice the cutset bound for sufficiently large $m$ and small $D$.
\item In the above proof, we considered only $m$ cuts. Can the bound be improved by considering more cuts? Based on our investigations, the answer appears to be negative. 
\item The above cutset lower bound can be readily extended to correlated WGN sources and weighted-sum computation. The resulting bound is tight for $m=2$ as shown in the previous section.
\end{enumerate}

\subsection{Upper Bound on $R^*(D)$ for Star Network}
\label{sec-star}
Consider a star network (or any network that contains a star network as a subnetwork) with $m$ nodes and $m-1$ edges $\Ec=\{\{1,2\},\{1,3\},\ldots,\{1,m\}\}$. For this network, we can establish the following upper bound on the network rate distortion function.

\medskip
\begin{proposition}
\label{prop-star}
The network rate distortion function for the star network is upper bounded by 
\begin{equation*}
R^*(D) \le \frac{m-1}{m}\log\left(\frac{2(m-1)^2}{m^3D}\right)
\end{equation*} 
for $D< (m-1)/m^2$.
\end{proposition}
\medskip

\begin{IEEEproof}
We use the following centralized protocol where node $1$ is treated as a ``cluster head.'' The centralized protocol has $T=2m-3$ rounds. The probability mass function $p(\ev)=1$ for the edge selection sequence $\ev=\{\{1,2\},\{1,3\},\ldots,\{1,m\},\{1,m-1\},\ldots,\{1,2\}\}$. There are two time slots in rounds $m-1$ and one time slot in the rest of the rounds. We first define the test channels
\[
\Xh_i = (1-d)(X_i+Z_i)
\]
for $i \in \Mc\setminus\{1\}$, where $(Z_2,Z_3,\ldots,Z_m)$ are independent WGN sources, each $Z_i$, $i \in \Mc\setminus\{1\}$ has average power $d/(1-d)$, and they are independent of $(X_1,X_2,\ldots,X_m)$. Define the sources
\[
S_1 = \frac{1}{m}X_1 + \frac{1}{m} \sum_{j \in \Mc\setminus\{1\}} \Xh_j,
\]
and
\[
U_i = \frac{1}{m}X_1 + \frac{1}{m} \sum_{j \in \Mc\setminus\{1,i\}} \Xh_j
\]
for $i \in \Mc\setminus\{1\}$. Since $S_1$ and $U_i$ for $i \in \Mc\setminus\{1\}$ are linear functions of WGN sources, they are also WGN sources. Now define the test channels 
\[
\Uh_i = (1-d_1)(U_i+\Zt_i)
\]
for $i \in \Mc\setminus\{1\}$, where $(\Zt_2,\Zt_3,\ldots,\Zt_m)$ are independent WGN sources, each $\Zt_i$, $i \in \Mc\setminus\{1\}$ has average power $\E(U_i^2)d_1/(1-d_1)$, and they are independent of $(X_1,X_2,\ldots,X_m)$ and $(Z_2,Z_3,\ldots,Z_m)$. Then $\Uh_i$ is Gaussian for $i \in \Mc\setminus\{1\}$. Define the sources
\[
S_i = \frac{1}{m}X_i + \Uh_i
\]
for $i \in \Mc\setminus\{1\}$, which are also Gaussian. We use these Gaussian test channels to generate Gaussian codes for $X_i$ and $U_i$ for every $i\in\Mc\setminus\{1\}$. These codes are revealed to all parties.

From the above test channels, we can readily compute the expected distortion between the average $S$ and the estimate $S_1$ for node 1 is 
\begin{align*}
\E\left((S-S_1)^2\right) &= \E\left(\left(\frac{1}{m}\sum_{i=2}^m (X_i-\Xh_i)\right)^2\right) \\
&= \frac{1}{m^2} \sum_{i=2}^m \E\left((X_i-\Xh_i)^2\right) = \frac{m-1}{m^2}d.
\end{align*}
The source $U_i$ has the following two properties
\begin{align*}
\E\left(\left(U_i-\frac{1}{m}\sum_{j\in\Mc\setminus\{i\}}X_j \right)^2\right) &= \E\left(\left(\frac{1}{m}\sum_{j\in\Mc\setminus\{1,i\}}(\Xh_j-X_j) \right)^2\right) \\
&= \frac{1}{m^2}\sum_{j\in\Mc\setminus\{1,i\}}\E\left((\Xh_j-X_j)^2\right) = \frac{m-2}{m^2}d
\end{align*}
and
\begin{align*}
\E\left(U_i^2 \right) 
&= \E\left(\left(\frac{1}{m}X_1 + \frac{1}{m}\sum_{j\in\Mc\setminus\{1,i\}}\Xh_j\right)^2\right) \\
&= \frac{1}{m^2} + \frac{1}{m^2}\sum_{j\in\Mc\setminus\{1,i\}}\E\left(\Xh_j^2\right) \\
&= \frac{1}{m^2} + \frac{m-2}{m^2}(1-d)
\end{align*}
for $i \in \Mc\setminus\{1\}$. The distortion for node $i\in\Mc\setminus\{1\}$ is 
\begin{align*}
\E\left((S - S_i)^2\right) &= \E\left(\left(\Uh_i-U_i + U_i-\frac{1}{m}\sum_{j\in\Mc\setminus\{i\}}X_j \right)^2\right) \\
&= \left( \E\left((\Uh_i-U_i)^2\right) + \E\left(\left( U_i-\frac{1}{m}\sum_{j\in\Mc\setminus\{i\}}X_j \right)^2\right) \right) \\
&= \left(\frac{1}{m^2}+\frac{m-2}{m^2}(1-d)\right)d_1 + \frac{m-2}{m^2}d.
\end{align*} 

The $(\ev,R(\ev),n)$ block code is specified as follows. In round $t\in[1:m-1]$, node $i=t+1$ uses the  Gaussian codes for $X_i$  to describe it to node $1$. Node $1$ then computes the estimates
\[
s_{11}^n = \frac{1}{m}x_{11}^n + \frac{1}{m} \sum_{j \in \Mc\setminus\{1\}} \xh_{j1}^n
\]
and
\[
u_{i1}^n = \frac{1}{m}x_{11}^n + \frac{1}{m} \sum_{j \in \Mc\setminus\{1,i\}} \xh_{j1}^n
\]
for $i\in\Mc\setminus\{1\}$, where $\xh_{j1}^n$ is the description of $x_{j1}^n$. In round $t\in[m-1:2m-3]$, node 1 uses the Gaussian code for the source $U_{2m-t-1}$ to describe it to node $2m-t-1$. At the end of round $2m-3$, node $i\in\Mc\setminus\{1\}$ computes the estimates 
\[
s_{i1}^n = \frac{1}{m}x_{i1}^n + \uh_{i1}^n,
\]
where $\uh_{i1}^n$ is the description of $u_{i1}^n$. 

In Theorem~\ref{thm-cutset}, we already showed that distortion $D\ge(m-1)/m^2$ is achievable using zero rate. Thus, we consider achievability for distortion $D<(m-1)/m^2$. 

Using a slight variation on Lemma~\ref{lm-achv}, we can show that the distortion $\E(X_i^2) d$ between $X_i$ and $\Xh_i$ and the distortion $\E(U_i^2) d_1$ between $U_i$ and $\Uh_i$ are achievable if 
\begin{align*}
r_i(i-1) &\ge I(X_i;\Xh_i) = \frac{1}{2}\log\frac{1}{d}, \\
r_1(2m-i-1) &\ge I(U_i; \Uh_i) = \frac{1}{2}\log\frac{1}{d_1}
\end{align*}
for $i \in \Mc\setminus\{1\}$, $0 < d < 1$ and $0 < d_1 < 1$.

Thus the per-node transmission rate is 
\begin{align*}
R &= \frac{1}{m}\left(\frac{m-1}{2}\log\frac{1}{d}+\frac{m-1}{2}\log\frac{1}{d_1}\right) 
= \frac{m-1}{2m}\log\left(\frac{1}{dd_1}\right),
\end{align*}
and we obtain the upper bound of the network rate distortion function, 
\begin{equation*}
R^*(D) \le \inf\left( \frac{m-1}{2}\log\left(\frac{1}{dd_1}\right) \right),
\end{equation*}
where the infimum is over all $0<d<1$ and $0<d_1<1$ satisfying
\begin{align*}
\frac{1}{m} \left( \frac{m-1}{m^2}d + (m-1)\left(\frac{1}{m^2}d_1+\frac{m-2}{m^2}(1-d)d_1 + \frac{m-2}{m^2}d \right)\right) &\le D 
\end{align*}
Choose $d=d_1=m^3D/(2(m-1)^2)$ 
satisfying the above constraint, we obtain the upper bound
\begin{equation*}
R^*(D) \le \frac{m-1}{m}\log\left(\frac{2(m-1)^2}{m^3D}\right).
\end{equation*}
\end{IEEEproof}
\medskip

Note that the above upper bound on $R^*(D)$ is not convex and therefore can be improved by time-sharing between the above centralized protocol and the degenerate zero-rate protocol. Note also that the ratio of the upper bound to the lower bound for $D < 1/m^2$ as $m\to \infty$ is less than or equal to 2. Thus a centralized protocol can achieve a rate within a factor of 2 of the cutset bound. 
\medskip

\noindent{\em Remark}: Note that this centralized protocol uses the minimum number of rounds $T=2m-3$. This does not imply optimality in terms of rate, however.

Can the factor of 2 between the upper bound and the cutset bound be tightened? It turns out that the cutset bound can be improved for low distortion for trees in general, which is illustrated in the following.
\medskip
 
\begin{proposition} 
\label{prop-star-cutset}
The network rate distortion function when the network is a tree is lower bounded by 
\[
R^*(D) \ge \frac{m-1}{2m}\log\left(\frac{1}{2m^3D^2}\right).
\]
\end{proposition}
\medskip

\begin{IEEEproof}
In a network with tree topology, removing an edge separates the network into two disconnected subnetworks. Fix $T$ and the edge selection sequence $\ev\in\Ec_1^T$, and consider an $(\ev,R(\ev),n)$ block code that achieves distortion $D(\ev)$. The total transmission rate $mR(\ev)$ is the number of bits flowing through each edge in both directions. Let $W_{ij}$ be the collection of indices sent from node $i$ to node $j$ and $R_{ij}(\ev)$ be the transmission rate for sending this message. Without loss of generality, assume that $i=1$, $j=m$, and removing the edge $\{1,m\}$ partitions the network $(\Mc,\Ec)$ into $(\Mc_1,\Ec_1)$ and $(\Mc_2,\Ec_2)$ such that $\Mc_1=\{1,2,\ldots,l\}$, $\Mc_2=\{l+1,l+2,\ldots,m\}$, and $l \ge l_0 = \lceil m/2 \rceil$. We first bound the number of bits flowing from node $1$ to node $m$.
\begin{align*}
nR_{1m}(\ev) &\ge H(W_{1m}) \ge H(W_{1m}| X_{l_0+1,1}^n, X_{l_0+2,1}^n, \ldots, X_{m1}^n) \\
&\ge I\left(\left. \frac{1}{m} \sum_{i=1}^{l_0} X_{i1}^n; W_{1m} \right| X_{l_0+1,1}^n, X_{l_0+2,1}^n, \ldots, X_{m1}^n \right) \\
&= \sum_{k=1}^n \left( h\left( \frac{1}{m}\sum_{i=1}^{l_0} X_{ik} \right) - h\left(\left. \frac{1}{m}\sum_{i=1}^{l_0} X_{ik} \right| W_{1m}, \frac{1}{m} \sum_{i=1}^{l_0} X_{i1}^{k-1}, X_{l_0+1,1}^n, X_{l_0+2,1}^n, \ldots, X_{m1}^n, \Sh_{mk} \right) \right) \\
&\ge \sum_{k=1}^n \left( h\left( \frac{1}{m}\sum_{i=1}^{l_0} X_{ik} \right) - h\left( S_k - \Sh_{mk} \right)  \right) \\
&\ge \frac{1}{2}\log\left(\frac{l_0}{m^2D_m}\right),
\end{align*}
where $D_m=(1/n)\sum_{k=1}^n \E((S_m-S_{mk})^2)$, the equality follows from the fact that the sources are independent WGN processes and $\Sh_{mk}$ is a function of $(W_{1m},X_{l+1,1}^n,X_{l+2,1}^n,\ldots,X_{m1}^n)$, and the last inequality follows by Jensen's inequality. Next we bound the number of bits flowing from node $m$ to node $1$. Consider
\begin{align*}
nR_{m1}(\ev) &\ge H(W_{m1}) \ge H(W_{m1}| X_{11}^n, X_{21}^n, \ldots, X_{m-1,1}^n) \\
&\ge I\left(\left. \frac{1}{m} X_{m1}^n; W_{m1} \right| X_{11}^n, X_{21}^n, \ldots, X_{m-1,1}^n \right) \\
&= \sum_{k=1}^n \left( h\left( \frac{1}{m} X_{mk} \right) - h\left(\left. \frac{1}{m} X_{mk} \right| W_{m1}, \frac{1}{m} X_{m1}^{k-1}, X_{11}^n, X_{21}^n, \ldots, X_{m-1,1}^n, \Sh_{1k} \right) \right) \\
&\ge \sum_{k=1}^n \left( h\left( \frac{1}{m} X_{mk} \right) - h\left( S_k - \Sh_{1k} \right)  \right) \\
&\ge \frac{1}{2}\log\left(\frac{1}{m^2D_1}\right),
\end{align*}
where $D_1=(1/n)\sum_{k=1}^n \E((S_1-S_{1k})^2)$, the equality follows from the fact that the source $X_m$ is a WGN process and $\Sh_{1k}$ is a function of $(W_{m1},X_{11}^n,$ $X_{21}^n, \ldots,X_{l1}^n)$, and the last inequality follows from Jensen's inequality. For any probability mass function $p(\ev)$ such that $\sum_{\ev\in\Ec_1^T} p(\ev)D(\ev) \le D$ where $D(\ev) = (1/m)\sum_{i=1}^m D_i$, the expected per-node transmission rate is lower bounded by
\[
\sum_{\ev \in \Ec_1^T} \frac{m-1}{m}p(\ev)(R_{1m}(\ev)+R_{m1}(\ev)) \ge \frac{m-1}{2m}\log\left(\frac{l_0}{m^4D^2}\right) 
\ge \frac{m-1}{2m}\log\left(\frac{1}{2m^3D^2}\right)
\]
for $D < 1/m^2$, where the first inequality follows from Jensen's inequality.
\end{IEEEproof}

\medskip
Note that as $D \to 0$, the ratio of the upper bound in Proposition~\ref{prop-star} to this lower bound approaches 1. The technique we use to tighten the cutset lower bound for this case, however, cannot be applied to networks with loops.

\section{Distributed Weighted-Sum Protocols}
\label{sec-gossip}
Again assume that the sources $(X_1,X_2,\ldots,X_m)$ are independent WGN processes each with average power one. We consider {\em distributed weighted-sum protocols} $(T,p(\ev),R,n,d)$ as defined in Section~\ref{sec-prelim}. The weighted-sum network rate distortion function $R^*_\mathrm{WS}(D)$ is difficult to establish in general. In the following subsection, we establish a lower bound on $R^*_\mathrm{WS}(D)$. In Subsection~\ref{sec-gossip-upper}, we establish upper and lower bounds on $R^*_\mathrm{GWS}(D)$ for gossip-based weighted-sum protocols, which in turn establishes an upper bound on $R^*_\mathrm{WS}(D)$. 

\subsection{Lower Bound on $R^*_\mathrm{WS}(D)$}
\label{sec-dist-lower}

We establish a lower bound on $R^*_\mathrm{WS}(D)$ that applies to any network. Consider a distributed weighted-sum protocol $(T,p(\ev),R,n,d)$ for a given network. Fix an edge selection sequence $\ev$ and let $t_{i\tau}$ be the $\tau$-th time node $i$ is selected and define
\[
\Tc_i = \{t_{i1},t_{i2},\ldots,t_{i T_i}\} = \{t: i \in e_t \}, \text{ for }i=1,2,\ldots,m,
\]
where  $T_i = |\Tc_i|$ is the number of rounds node in which $i$ is selected. Then the number of rounds $T$ can be expressed as $T = (1/2)\sum_{i=1}^m T_i$, where the factor of $1/2$ is due to the fact that two nodes are selected in each round. We shall need the following properties of the estimate $S_i(t)$ to prove the lower bound.

\medskip
\begin{lemma}
\label{lm-dist-gamma}
For any distributed weighted-sum protocol $(T,p(\ev),R,n,d)$ and any edge selection sequence $\ev\in\Ec_1^T$, the estimate of node $i$ at the end of round $t$ can be expressed as
\begin{equation}
\label{equ-cvx-comb}
S_i(t) = \sum_{j=1}^m \gamma_{ij}(t)S_j(0) + V_i(t),
\end{equation}
where $V_i(t)$ is Gaussian and independent of the sources $(X_1,X_2,\ldots,X_m)$. Furthermore, the diagonal coefficients satisfy the following property
\begin{equation}
\label{equ-gamma}
\gamma_{ii}(t) \ge \frac{1}{2^\tau} 
\end{equation}
for $t_{i\tau} \le t < t_{i,\tau+1}$ and $\tau \in [1:T_i]$.
\end{lemma}
\medskip

\begin{IEEEproof}
When $t=0$, we have $\gamma_{ii}(0)=1$ and $\gamma_{ij}(0)=0$ for $i\ne j$. Suppose that $\gamma_{ij}(t)\ge 0$ for $i,j \in \Mc$ and (\ref{equ-cvx-comb}) holds up to round $t$. In round $t+1$, assume that the node pair $\{i,l\}$ is selected. By the update equations in~\eqref{equ-gossip-s}, the estimate of node $i$ at the end of this round is
\begin{align*}
S_i(t+1) &= \frac{1}{2} S_i(t) + \frac{1}{2(1-d)} \Sh_l(t) \\
&= \frac{1}{2} S_i(t) + \frac{1}{2} S_l(t) + \frac{1}{2} Z_l(t) \\
&= \sum_{j=1}^m \left(\frac{1}{2}\gamma_{ij}(t)+\frac{1}{2}\gamma_{lj}(t)\right)S_j(0) + \frac{1}{2} V_i(t) + \frac{1}{2} V_l(t) + \frac{1}{2} Z_l(t),
\end{align*}
where $Z_l(t)$ is a WGN independent of $(X_1, X_2, \ldots, X_m)$. Thus $V_i(t+1)=(V_i(t) + V_l(t) + Z_l(t))/2$ is independent of the sources $(X_1, X_2, \ldots, X_m)$, and the coefficient $\gamma_{ij}(t+1)=(\gamma_{ij}(t)+\gamma_{lj}(t))/2 \ge 0$. By induction, (\ref{equ-cvx-comb}) and $\gamma_{ij}(t)\ge 0$ hold for all $t \in [1:T]$. Therefore, 
\begin{align*}
\gamma_{ii}(t+1) = \frac{1}{2}\gamma_{ii}(t) + \frac{1}{2}\gamma_{li}(t) \ge \frac{1}{2}\gamma_{ii}(t). 
\end{align*}
This can be rewritten as 
\begin{align*}
\gamma_{ii}(t+1) &\ge
\begin{cases}
\frac{1}{2}\gamma_{ii}(t) & \text{if}\ i\in e_{t+1} \\[4pt]
\gamma_{ii}(t) & \text{if}\ i\notin e_{t+1},
\end{cases} 
\end{align*}
and we have (\ref{equ-gamma}).
\end{IEEEproof}
\medskip

Using this lemma, we establish the following lower bound on the number of rounds $T$ for any distributed weighted-sum protocol and any network.

\medskip
\begin{lemma}
\label{lm-dist-round}
Given $0<D<(m-1)/m^2$, if a distributed weighted-sum protocol $(T,p(\ev),R,n,d)$ achieves distortion $D$, then
\[
T \ge \frac{m}{2}\log\left(\frac{1}{\sqrt{D}+1/m}\right).
\]
\end{lemma}

\medskip

\begin{IEEEproof}
By Lemma~\ref{lm-dist-gamma}, given any edge selection sequence $\ev \in \Ec_1^T$, the estimate of node $i$ at round $T$ is
\begin{align*}
S_i(T) = \sum_{j=1}^m \gamma_{ij}(T)S_j(0) + V_i(T).
\end{align*}
The distortion at node $i$ is
\begin{align*}
\E\left((S-S_i(T))^2\right) &\ge \left(\gamma_{ii}(T)-\frac{1}{m}\right)^2 \ge \left(\frac{1}{2^{T_i}}-\frac{1}{m}\right)^2.
\end{align*}
If $(\ev,R(\ev),n)$ block code achieves distortion $D(\ev)$, then a lower bound on the number of rounds can be found by solving the optimization problem
\begin{align*}
\text{minimize }\hspace{4pt} & \frac{1}{2}\sum_{i=1}^m T_i \\
\text{subject to }\hspace{4pt} & \frac{1}{m}\sum_{i=1}^m\left(\frac{1}{2^{T_i}}-\frac{1}{m}\right)^2 \le D(\ev) \\
& T_i \ge 0\ \text{for}\ i\in\Mc,
\end{align*}
where $T_i$ is real-valued for $i \in \Mc$. Let $y_i = 1/2^{T_i} \le 1$. The above optimization problem reduces to the convex optimization problem 
\begin{align*}
\text{minimize }\hspace{4pt} & - \frac{1}{2}\sum_{i=1}^m \log y_i \\
\text{subject to }\hspace{4pt} & \frac{1}{m}\sum_{i=1}^m\left(y_i-\frac{1}{m}\right)^2 \le D(\ev) \\
& y_i \le 1\ \text{for}\ i\in\Mc,
\end{align*}
and the KKT conditions are necessary and sufficient for optimality. The Lagrangian for this problem is
\[
L(y_1^m,\nu,\mu_1^m) = -\frac{1}{2}\sum_{i=1}^m \log y_i + \nu\left(\frac{1}{m}\sum_{i=1}^m\left(y_i-\frac{1}{m}\right)^2 - D(\ev)\right) + \mu_i(y_i-1)
\]
for $\nu \ge 0$ and $\mu_i \ge 0$, $i \in \Mc$. Setting $\partial L/\partial y_i =0$, we have $y_i=y_j$ for $i\ne j$. Thus, 
\begin{align*}
T_i = -\log y_i &\ge \log\left(\frac{1}{\sqrt{D(\ev)}+1/m}\right), 
\end{align*}
and the minimum number of rounds is lower bounded by
$T = (1/2)\sum_{i=1}^m T_i \ge (m/2)\log(1/(\sqrt{D(\ev)}+1/m))$. For any distributed weighted-sum protocol that achieves distortion $D$, there exists at least one $(\ev,R(\ev),n)$ block code that achieves distortion $D(\ev) \le D$. Thus, $T \ge (m/2)\log(1/(\sqrt{D}+1/m))$.
\end{IEEEproof}
\medskip

We are now ready to establish the lower bound on  $R^*_\mathrm{WS}(D)$.

\medskip
\begin{theorem}
\label{thm-dist-lower}
Given $0<D<(m-1)/m^2$, then 
\[
R^*_\mathrm{WS}(D) \ge \frac{1}{2}\left(\log\frac{1}{\sqrt{D}+1/m}\right)\left(\log\frac{1}{4mD}\right).
\]
\end{theorem}

\medskip
\begin{IEEEproof}
Given a distributed weighted-sum protocol $(T,p(\ev),R,n,d)$. Fix an edge selection sequence $\ev\in\Ec_1^T$. Suppose that the edge selected at round $t_{i\tau}$ is $\{i,j\}$. Then at the end of this round, the estimate for node $j$ is 
\begin{align*}
S_j(t_{i\tau}) &= \frac{1}{2} S_j(t_{i\tau}-1) + \frac{1}{2} S_i(t_{i\tau}-1) + \frac{1}{2} Z_i(t_{i\tau}-1),
\end{align*}
where $Z_i(t_{i\tau}-1)$ is a WGN with average power $\E\left(S_i(t_{i\tau}-1)^2\right)d/(1-d)$. By induction, we can show that the estimate of node $l$ at time $t\ge t_{i\tau}$ has the form $S_l(t) = (1/2)\beta_l(t) Z_i(t_{i\tau}-1) + \St_l(t)$, where $\beta_l(t)\ge 0$, $\sum_{l=1}^m\beta_l(t)=1$, and $\St_l(t)$ is independent of $Z_i(t_{i\tau}-1)$. Now we compute the average distortion at the end of round $T$
\begin{align*}
\frac{1}{m}\sum_{l=1}^m \E\left((S-S_l(T))^2\right) &= \frac{1}{m}\sum_{l=1}^m\left( \E\left(\left(\frac{\beta_l(T)}{2} Z_i(t_{i\tau}-1)\right)^2\right) + \E\left((S-\St_l(T))^2\right) \right) \\
&\ge \frac{1}{m^2}\E\left(\left(\frac{1}{2} Z_i(t_{i\tau}-1)\right)^2\right) + \frac{1}{m}\sum_{l=1}^m\E\left((S-\St_l(T))^2\right) \\
&\ge \frac{d}{4m^2(1-d)}\gamma_{ii}(t_{i\tau}-1)^2 + \frac{1}{m}\sum_{l=1}^m\E\left((S-\St_l(T))^2\right) \\
&\ge \frac{d}{4m^2(1-d)2^{2(\tau-1)}} + \frac{1}{m}\sum_{l=1}^m\E\left((S-\St_l(T))^2\right), 
\end{align*}
where the first inequality follows by the Cauchy--Schwarz inequality, and the last inequality follows from Lemma~\ref{lm-dist-gamma}. We can repeat the above arguments for the second term $(1/m)\sum_{l=1}^m\E\left((S-\St_l(T))^2\right)$ and we obtain
\begin{align*}
\frac{1}{m}\sum_{l=1}^m \E\left(S-S_l(T))^2\right) &\ge \sum_{l=1}^m\sum_{\tau=1}^{T_l}\frac{d}{4m^2(1-d)2^{2(\tau-1)}} \ge \frac{d}{4m}.
\end{align*}

Since at least one $(\ev,R(\ev),n)$ block code has distortion $D(\ev)\le D$, the normalized distortion is upper bounded by $d\le 4mD$. Thus, by Lemma~\ref{lm-achv}, the average rate is lower bounded by
\[
R = \frac{T}{m}\log\frac{1}{d} \ge \frac{1}{2}\left(\log\frac{1}{\sqrt{D}+1/m}\right)\left(\log\frac{1}{4mD}\right).
\]
This completes the proof of the theorem.
\end{IEEEproof}
\medskip

\noindent{\em Remark}: The above lower bound and the cutset bound in Theorem~\ref{thm-cutset} differ in order by roughly a factor of $\log m$, since $\log\left(1/(\sqrt{D}+1/m)\right)$ is on the order of $\log m$ for all $D<(m-1)/m^2$. Given that a centralized protocol for the star network can come to within a factor of 2 of the cutset bound suggests that the $\log m$ factor is the penalty of using distributed versus centralized protocols. 

\subsection{Bounds on $R^*_\mathrm{GWS}(D)$}
\label{sec-gossip-upper}
In this section, we establish bounds on $R^*_\mathrm{GWS}(D)$ for {\em gossip-based weighted-sum protocols} $(T,Q,R,n,d)$ defined in Section~\ref{sec-prelim}. Note that this result also establishes an upper bound on $R^*(D)$ and $R^*_\mathrm{WS}(D)$ because $R^*(D) \le R^*_\mathrm{WS}(D) \le R^*_\mathrm{GWS}(D)$. 

Let $\Sv(t)=[S_1(t)\ S_2(t)\ \ldots\ S_m(t)]^T$ and rewrite the update equations (\ref{equ-gossip-s}) in a matrix form as
\begin{equation}
\label{equ-gossip-sys}
\Sv(t+1) = A(t+1)\Sv(t) + \Zv(t+1), 
\end{equation}
where (i) $A(t+1)$ is an $m \times m$ random matrix such that
\begin{align*}
A(t+1) &= A_{ij} = I - \frac{1}{2}(\phiv_i-\phiv_j)(\phiv_i-\phiv_j)^T 
\end{align*}
with probability $(1/m)Q_{ij}$, independent of $t$, where $I$ is the identity matrix and $\phiv_i$ and $\phiv_j$ are unit vectors along the $i$-th and $j$-th axes, and (ii) 
\begin{align*}
\Zv(t+1) &= \frac{1}{2}Z_j(t)\phiv_i + \frac{1}{2}Z_i(t)\phiv_j,
\end{align*}
where $Z_i(t)$ and $Z_j(t)$ are WGN sources with average power $\E\left(S_i(t)^2\right)d/(1-d)$ and $\E\left(S_j(t)^2\right)d/(1-d)$, respectively, defined in Section~\ref{sec-prelim}. 

Recall the following properties of the matrix $A(t)$ from~\cite{BoydGPS05}.
\begin{enumerate}
\item $\E\left(A(t)^TA(t)\right)=A$, where $A=\E(A(0))$ and the expectation is taken over all $A_{ij}$ with probability $(1/m)Q_{ij}$.
\item $A(t)$ is symmetric positive semidefinite.
\item The largest eigenvalue of $A$ is 1 and the corresponding eigenvector is $\onev=[1\ 1\ \ldots\ 1]^T$.
\item The stochastic matrix $Q$ that minimizes the second largest eigenvalue of $A$ is the solution to the  optimization problem
\begin{align}
\label{equ-gossip-eig}
\text{minimize }\hspace{4pt} & \lambda_2(A) \\
\text{subject to}\hspace{4pt} & A = \sum_{i=1}^m\sum_{j=1}^m \frac{1}{m}Q_{ij}A_{ij} \nonumber\\
& Q_{ij} \ge 0 \text{ for all } i,j \nonumber\\
& Q_{ij} = 0 \text{ if } \{i,j\}\notin\Ec \nonumber\\
& \sum_{j=1}^m Q_{ij}=1 \text{ for all } i. \nonumber
\end{align}
Let $\lambda_2$ be the second largest eigenvalue of the matrix $A$ associated with the optimum matrix $Q^*$, which is a function of the topology of the network.
\end{enumerate}

Referring to the linear dynamical system in (\ref{equ-gossip-sys}), express $\Sv(T)$ as
\begin{equation}
\label{equ-gossip-est}
\Sv(T) = A(T,1)\Sv(0) + \sum_{t=1}^T A(T,t+1)\Zv(t) = A(T,1)\Sv(0) + \Vv(T),
\end{equation}
where $\Vv(T)=\sum_{t=1}^T A(T,t+1)\Zv(t)$ and 
\[
A(t_2,t_1) = \begin{cases} 
A(t_2)A(t_2-1)\ldots A(t_1) &\text{ if } t_2\ge t_1 \\[4pt]
I &\text{ if } t_2<t_1.
\end{cases}
\]

We will need the following lower bound on the number of rounds $T$ to prove the lower bound.

\medskip
\begin{lemma}
\label{lm-gossip-round}
Given a connected network, if a gossip-based weighted-sum protocol $(T,Q,R,n,d)$ achieves distortion $D$, then
\[
T \ge \frac{m-1}{2}\ln\left(\frac{m-1}{mD}\right).
\]
\end{lemma}
\medskip

\begin{IEEEproof}
Assume that the matrix $A$ has eigenvalues
\[
\lambda_1=1\ge\lambda_2\ge\lambda_3\ge\ldots\ge\lambda_m
\]
with corresponding orthonormal eigenvectors $\av_1=(1/\sqrt{m})\onev,\av_2,\av_3,\ldots,\av_m$. We can express the estimate $\Sv(0)$ as $\Sv(0)=\sum_{i=1}^mS_i\av_i$, where $S_i=\av_i^T\Sv(0)\sim\Norm(0,1)$. 

Consider the sum of distortions over all nodes for an $(\ev,R(\ev),n)$ block code at the end of round $T$,
\begin{align*}
\E\left(\|\Sv(T)-J\Sv(0)\|^2\right) &= \E\left(\|A(T,1)\Sv(0)-J\Sv(0)\|^2\right) + \E\left(\|\Vv(T)\|^2\right) \\
&\ge \E\left( \E\left(\left. \|A(T,1)\Sv(0)-J\Sv(0)\|^2 \right|\Sv(0)\right) \right) \\
&\ge \E\left( \E\left(\left. A(T,1)\Sv(0)-J\Sv(0) \right|\Sv(0)\right)^T \E\left(\left. A(T,1)\Sv(0)-J\Sv(0) \right|\Sv(0)\right) \right) \\
&= \E\left( \|\E(A(T,1))\Sv(0)-J\Sv(0)\|^2 \right) \\
&= \E\left( \left\|\sum_{i=2}^m S_i\lambda_i^T\av_i\right\|^2 \right) \\
&= \sum_{i=2}^m \lambda_i^{2T} \ge (m-1)\left(\frac{1}{m-1}\sum_{i=2}^m\lambda_i\right)^{2T},
\end{align*}
where the second and last inequalities follow from Jensen's inequality.

By the definition of the matrix $A$, 
\begin{align*}
\sum_{i=1}^m \lambda_i &= \tr(A) = \tr\left(\sum_{i=1}^m\sum_{j=1}^m \frac{1}{m}Q_{ij} \left(I-\frac{1}{2}(\phiv_i-\phiv_j)(\phiv_i-\phiv_j)^T\right)\right) \\
&= \sum_{i=1}^m\sum_{j=1}^m \frac{1}{m}Q_{ij} (m-1) = m-1.
\end{align*}
Thus, $\sum_{i=2}^m\lambda_i = m-2$.

To achieve distortion $D(\ev)$, 
\begin{align*}
T &\ge \frac{\ln\left((m-1)/mD(\ev)\right)}{2\ln\left((m-1)/(m-2)\right)} 
\ge \frac{m-1}{2}\ln\left(\frac{m-1}{mD(\ev)}\right).
\end{align*}
Since at least one $(\ev,R(\ev),n)$ block code achieves distortion $D(\ev)\le D$,
\[
T \ge \frac{m-1}{2}\ln\left(\frac{m-1}{mD}\right).
\]
\end{IEEEproof}
\medskip

The following lemma is useful for calculating norm squared of vectors relating to $A(t)$.

\medskip
\begin{lemma}
\label{lm-gossip-mat}
For any random vector $\Yv$, independent of $A(t)$, we have
\begin{itemize}[\IEEEsetlabelwidth{(ii)}]
\item[(i)] $\E\left(\|A(t)\Yv\|^2\right) \le \lambda_2(A)\E\left(\|\Yv-J\Yv\|^2\right) + \E\left(\|J\Yv\|^2\right)$, and
\item[(ii)] $\E\left(\|A(t)\Yv-J\Yv\|^2\right) \le \lambda_2(A)\E\left(\|\Yv-J\Yv\|^2\right)$,
\end{itemize}
where $J =(1/m)\onev \onev^T$.
\end{lemma}
\medskip

The proof of the lemma is given in the Appendix.

The following lemma gives an upper bound on the average distortion for a gossip-based weighted-sum protocol.

\medskip
\begin{lemma}
\label{lm-gossip-dist}
The average per-letter distortion of the gossip-based weighted-sum protocol $(T,Q,R,n,d)$ is upper bounded by
\begin{align*}
\frac{1}{m}\E\left(\|\Sv(T)-J\Sv(0)\|^2\right) 
&\le \frac{1}{m^2}\left((1+u)^T-1\right) + \frac{1}{m}\frac{u}{1-\lambda_2+u}(1+u)^T + \frac{1}{m}\frac{u}{1-\lambda_2-u} + (\lambda_2+u)^T,
\end{align*}
where $u =d/2m(1-d)$.
\end{lemma}
\medskip

\begin{IEEEproof}
Consider the sum of distortions
\begin{align*}
\E\left(\|\Sv(T)-J\Sv(0)\|^2\right) &= \E\left(\|A(T,1)\Sv(0)-J\Sv(0)\|^2\right) + \E\left(\|\Zv(T)\|^2\right) \\
&= \E\left(\|A(T,1)\Sv(0)-J\Sv(0)\|^2\right) + \sum_{t=1}^T \E\left(\|A(T,t+1)\Wv(t)\|^2\right), 
\end{align*}
where the first term corresponds to the sum distortion for the infinite-rate gossip algorithm and the second term is contributed by quantization distortions.
We first find an upper bound on the first term using Lemma~\ref{lm-gossip-mat}. Consider
\begin{align*}
\E\left(\|A(T,1)\Sv(0)-J\Sv(0)\|^2\right) &= \E\left(\|A(T,1)\Sv(0)-JA(T-1,1)\Sv(0)\|^2\right) \\
&\le \lambda_2\E\left(\|A(T-1,1)\Sv(0)-JA(T-1,1)\Sv(0)\|^2\right) \\
&\le \lambda_2\E\left(\|A(T-1,1)\Sv(0)-J\Sv(0)\|^2\right) \\
&\le \lambda_2^T\E\left(\|\Sv(0)-J\Sv(0)\|^2\right) = (m-1)\lambda_2^T.
\end{align*}
Next we consider
\begin{align*}
& \E\left(\|A(T,t+1)\Zv(t)\|^2\right) \\
&\qquad \le \lambda_2\E\left(\|A(T-1,t+1)\Zv(t)-JA(T-1,t+1)\Zv(t)\|^2\right) + \E\left(\|JA(T-1,t+1)\Zv(t)\|^2\right) \\
&\qquad = \lambda_2\E\left(\|A(T-1,t+1)\Zv(t)-JA(T-2,t+1)\Zv(t)\|^2\right) + \E\left(\|J\Zv(t)\|^2\right) \\
&\qquad \le \lambda_2^2\E\left(\|A(T-2,t+1)\Zv(t)-JA(T-2,t+1)\Zv(t)\|^2\right) + \E\left(\|J\Zv(t)\|^2\right) \\
&\qquad \le \lambda_2^{T-t}\E\left(\|\Zv(t)-J\Zv(t)\|^2\right) + \E\left(\|J\Zv(t)\|^2\right) \\
&\qquad = \left(\frac{m-1}{m}\lambda_2^{T-t}+\frac{1}{m}\right)\E\left(\|\Zv(t)\|^2\right).
\end{align*}
By the definition of the vector $\Zv(t)$, we have
\begin{align*}
\E\left(\|\Zv(t)\|^2\right) &= \frac{d}{2m(1-d)}\E\left(\|\Sv(t-1)\|^2\right) \\
&= u\left(\E\left(\|(A(t-1,1)\Sv(0)\|^2\right) + \E\left(\|\Vv(t-1)\|^2\right) \right) \\
&\le u\left(\lambda_2\E\left(\|A(t-2,1)\Sv(0)-JA(t-2,1)\Sv(0)\|^2\right) + \E\left(\|JA(t-2,1)\Sv(0)\|^2\right) + \E\left(\|\Vv(t-1)\|^2\right) \right) \\
&\le u\left(\lambda_2^{t-1}\E\left(\|\Sv(0)-J\Sv(0)\|^2\right) + \E\left(\|J\Sv(0)\|^2\right) + \E\left(\|\Vv(t-1)\|^2\right)\right) \\
&\le u\left(\left(\frac{m-1}{m}\lambda_2^{t-1}+\frac{1}{m}\right)\E\left(\|\Sv(0)\|^2\right) + \E\left(\|\Vv(t-1)\|^2\right)\right) \\
&\le u\left( 1 + (m-1)\lambda_2^{t-1} + \E\left(\|\Vv(t-1)\|^2\right) \right). 
\end{align*}
Combining above inequalities, we obtain
\begin{align*}
\E\left(\|\Vv(t)\|^2\right) &\le \sum_{\tau=1}^t u\left(\frac{1}{m}+\frac{m-1}{m}\lambda_2^{t-\tau}\right)\left(1 + (m-1)\lambda_2^{\tau-1} + \E\left(\|\Vv(\tau-1)\|^2\right)\right).
\end{align*}
Suppose that for $\tau=1,2,\ldots,t$
\begin{equation}
\label{equ-ind-zv}
\E\left(\|\Vv(\tau)\|^2\right) \le (1+u)^{\tau}+(m-1)(\lambda_2+u)^{\tau}-1-(m-1)\lambda_2^{\tau},
\end{equation}
then 
\begin{align*}
\E\left(\|\Vv(t+1)\|^2\right) &\le \sum_{\tau=1}^{t+1} u\left(\frac{1}{m}+\frac{m-1}{m}\lambda_2^{t-\tau+1}\right)\left((1+u)^{\tau-1}+(m-1)(\lambda_2+u)^{\tau-1}\right) \\
&\le \sum_{\tau=1}^{t+1} u\left((1+u)^{\tau-1}+(m-1)\left(\frac{\lambda_2+u}{\lambda_2}\right)^{\tau-1}\lambda_2^t\right) \\
&= u\frac{(1+u)^{t+1}-1}{u} +u(m-1)\frac{(\lambda_2+u)^{t+1}-(\lambda_2)^{t+1}}{u} \\
&= (1+u)^{t+1}+(m-1)(\lambda_2+u)^{t+1}-1-(m-1)\lambda_2^{t+1}.
\end{align*}
By induction, (\ref{equ-ind-zv}) holds for $\tau=1,2,\ldots,T-1$, thus
\begin{align*}
\E\left(\|\Vv(T)\|^2\right) &\le \sum_{t=1}^T u\left(\frac{1}{m}+\frac{m-1}{m}\lambda_2^{T-t}\right)\left((1+u)^t+(m-1)(\lambda_2+u)^t\right) \\
&= \frac{1}{m}\left((1+u)^T-1\right)+\frac{m-1}{m}\frac{u}{1-\lambda_2+u}\left((1+u)^T-\lambda_2^T\right) \\
&\qquad +\frac{m-1}{m}\frac{u}{1-\lambda_2-u}\left(1-(\lambda_2+u)^T\right)+\frac{(m-1)^2}{m}\left((\lambda_2+u)^T-\lambda_2^T\right).
\end{align*}
Therefore, we have the upper bound on distortion
\begin{align*}
& \frac{1}{m}\E\left(\|\Sv(T)-J\Sv(0)\|^2\right) \\
&\qquad \le \frac{m-1}{m}\lambda_2^T + \frac{1}{m^2}\left((1+u)^T-1\right)+\frac{m-1}{m^2}\frac{u}{1-\lambda_2+u}\left((1+u)^T-\lambda_2^T\right) \\
&\qquad + \frac{m-1}{m^2}\frac{u}{1-\lambda_2-u}\left(1-(\lambda_2+u)^T\right) + \left(\frac{m-1}{m}\right)^2\left((\lambda_2+u)^T-\lambda_2^T\right) \\
&\qquad \le \frac{1}{m^2}\left((1+u)^T-1\right) + \frac{1}{m}\frac{u}{1-\lambda_2+u}(1+u)^T + \frac{1}{m}\frac{u}{1-\lambda_2-u} + (\lambda_2+u)^T.
\end{align*}
\end{IEEEproof}
\medskip

We are now ready to establish the bounds on $R^*_\mathrm{GWS}(D)$. 

\medskip
\begin{theorem}
\label{thm-dist-upper}
Given a connected network with associated eigenvalue $\lambda_2$, then  
\begin{itemize}[\IEEEsetlabelwidth{(ii)}]
\item[(i)] if a gossip-based weighted-sum protocol achieves distortion $D<1/4m$, then
\[
R^*_\mathrm{GWS}(D) \ge \frac{m-1}{2m}\left(\ln\frac{m-1}{mD}\right)\left(\log\frac{1}{4mD}\right), \text{ and }
\]
\item[(ii)] there exists an $m(D)$ and a gossip-based weighted-sum protocol such that for all $m\ge m(D)$,
\[
R^*_\mathrm{GWS}(D) \le \frac{1}{m\bar\lambda_2}\left(\ln\frac{2}{D}\right)\left(\log\frac{\ln(2/D)}{m^2\bar\lambda_2D}\right), 
\]
where $\bar\lambda_2=1-\lambda_2$.
\end{itemize}
\end{theorem}
\medskip

\begin{IEEEproof}
\begin{itemize}[\IEEEsetlabelwidth{(ii)}]
\item[(i)] The distortion analysis is the same as Theorem~\ref{thm-dist-lower}. Thus, $d/4m \le D$. Using Lemma~\ref{lm-achv} and~\ref{lm-gossip-round}, we have the following lower bound on the expected per-node transmission rate
\begin{align*}
R &= \frac{T}{m}\log\left(\frac{1}{d}\right) \ge \frac{m-1}{2m}\left(\ln\frac{m-1}{mD}\right)\left(\log\frac{1}{4mD}\right). 
\end{align*}
\item[(ii)] If the distortion $D\ge (m-1)/m^2$, then zero rate is achievable for $T=0$ as discussed in Section~\ref{sec-cutset}. Otherwise, we choose the optimal stochastic matrix $Q^*$ with eigenvalue $\lambda_2$ according to (\ref{equ-gossip-eig}) for the given network topology. 

For $D< (m-1)/m^2, $ we need to show that 
\[
\lim_{m\to\infty} \frac{1}{mD}\E\left(\|\Sv(T)-J\Sv(0)\|^2\right) < 1.
\]
By Lemma~\ref{lm-gossip-dist}, it suffices to show that
\[
\lim_{m\to\infty} \left( \frac{1}{m^2D}\left((1+u)^T-1\right)+\frac{1}{mD}\frac{u}{\bar\lambda_2+u}(1+u)^T+\frac{1}{mD}\frac{u}{\bar\lambda_2-u}+\frac{1}{D}(1-\bar\lambda_2+u)^T \right) < 1.
\]
We set
\begin{align*}
T = \frac{1}{\bar\lambda_2}\ln\left(\frac{2}{D}\right)\ \text{and}\ 
d = \frac{m^2\bar\lambda_2D}{\ln(2/D)}.
\end{align*}
First consider
\begin{align*}
\lim_{m\to\infty} \frac{1}{mD}\left((1+u)^T-1\right) &= \lim_{m\to\infty} \frac{1}{mD}\left(\left(1+\frac{m\bar\lambda_2D}{2(1-d)\ln(2/D)}\right)^{(1/\bar\lambda_2)\ln(2/D)}-1\right) \\
&= \begin{cases}
\displaystyle \frac{1}{c}\left(e^{c/2}-1\right) &\displaystyle  \text{ if } \lim_{m\to\infty} mD = c \\[8pt]
\displaystyle \frac{1}{2} &\displaystyle \text{ if } \lim_{m\to\infty} mD = 0,
\end{cases}\\
\lim_{m\to\infty} (1+u)^T &= \lim_{m\to\infty} \left(1+\frac{m\bar\lambda_2D}{2(1-d)\ln(2/D)}\right)^{(1/\bar\lambda_2)\ln(2/D)} < e^{1/2}, \\
\lim_{m\to\infty} \frac{1}{mD}\left(\frac{u}{\bar\lambda_2\pm u}\right) &= \lim_{m\to\infty} \frac{1}{mD}\left(\frac{m\bar\lambda_2D/2(1-d)\ln(2/D)}{\bar\lambda_2\pm m\bar\lambda_2D/2(1-d)\ln(2/D)}\right) = 0, \\
\lim_{m\to\infty} \frac{1}{D}(1-\bar\lambda_2+u)^T &= \lim_{m\to\infty} \frac{1}{D}\left(1-\bar\lambda_2+\frac{m\bar\lambda_2D}{2(1-d)\ln(2/D)}\right)^{(1/\bar\lambda_2)\ln(2/D)} < \frac{e^{1/2}}{2}.
\end{align*}
Letting $m\to\infty$, we obtain the following ratio of the expected distortion to the given distortion
\[
\lim_{m\to\infty}\frac{1}{mD}\E\left(\|\Sv(T)-J\Sv(0)\|^2\right) \le \frac{e^{1/2}}{2} < 1.
\]
Therefore, by Lemma~\ref{lm-achv}, the average distortion $D$ is achievable for average rate
\[
R = \frac{T}{m}\log\frac{1}{d} = \frac{1}{m\bar\lambda_2}\left(\ln\frac{2}{D}\right)\left(\log\frac{\ln(2/D)}{m^2\bar\lambda_2D}\right). 
\]
\end{itemize}
\end{IEEEproof}
Note that the upper bound on $R^*_\text{GWS}(D)$ is not in general convex and therefore can be improved by time-sharing.
\medskip

\noindent{\em Remark}: 
The above upper bound can be improved for distortion decreasing slowly in $m$ by choosing higher $d$. Specifically, by choosing $T=(1/\bar\lambda_2)\ln(3/D)$ and $d=2m^2\bar\lambda_2D/\ln m$, for distortion $D=\Omega(1/m\log m)$, and  $T=(1/\bar\lambda_2)\ln(3/D)$ and $d=m^2\bar\lambda_2D/4$ for distortion $D=o(1/m\log m)$ and $D=\Omega(m^{-c})$, $c>0$, we obtain the following tighter upper bounds
\[
R^*_\mathrm{GWS}(D) \le \begin{cases}
\displaystyle \frac{1}{m\bar\lambda_2}\left(\ln\frac{3}{D}\right)\left(\log\frac{\ln m}{2m^2\bar\lambda_2D}\right) &\displaystyle \text{ if } D=\Omega\left(\frac{1}{m\log m}\right) \\[8pt]
\displaystyle \frac{1}{m\bar\lambda_2}\left(\ln\frac{8}{D}\right)\left(\log\frac{4}{m^2\bar\lambda_2D}\right) &\displaystyle  \text{ if } D=o\left(\frac{1}{m\log m}\right) \text{ and } D=\Omega\left(m^{-c}\right).
\end{cases}
\]
Note that these upper bounds differ from the upper bound given in Theorem~\ref{thm-dist-upper} but have the same order.
\medskip

Since Gaussian sources are the hardest to compress, we can show that the above upper bound is also an upper bound for general, non-Gaussian sources.

\medskip
\begin{corollary}
The upper bound in Theorem~\ref{thm-dist-upper} holds for general non-Gaussian i.i.d. sources $(X_1,X_2,\ldots,X_m)$ each with average power one. 
\end{corollary}
\medskip

\begin{IEEEproof}
Assume that Gaussian test channels with independent additive WGN are used. Then the analysis of distortion is the same as in the proof of Theorem~\ref{thm-dist-upper}. In round $t+1$, the rate of node $i\in e_{t+1}$ is
\begin{align*}
I(S_i(t);\Sh_i(t)) &= h(\Sh_i(t)) - h(\Sh_i(t)|S_i(t)) \\
&= h\left((1-d)(S_i(t)+Z_i(t))\right) - h((1-d)Z_i(t)) \\
&= h\left((1-d)(S_i(t)+Z_i(t))\right) - \frac{1}{2}\log\left(2\pi e \E\left(S_i(t)^2\right) d(1-d) \right) \\
&\le \frac{1}{2}\log\left(2\pi e \E(S_i(t)^2) (1-d) \right) - \frac{1}{2}\log\left(2\pi e \E\left(S_i(t)^2\right) d(1-d) \right) \\
&= \frac{1}{2}\log\frac{1}{d},
\end{align*}
which is less than the rate for Gaussian sources. Thus, the upper bound on the expected network rate distortion function for WGN sources in Theorem~\ref{thm-dist-upper} is also an upper bound for non-Gaussian i.i.d. sources with the same average powers.
\end{IEEEproof}
\medskip

\noindent{\em Remarks}:
\begin{enumerate}
\item For a complete graph, it can be easily shown that $\lambda_2 = 1-1/(m-1)$. 
Thus the upper and lower bounds of Theorem~\ref{thm-dist-upper} gives
\[
R^*_\mathrm{GWS}(D) \le \frac{m-1}{m}\left(\ln\frac{2}{D}\right)\left(\log\frac{(m-1)\ln(2/D)}{m^2D}\right)
\]
and
\[
R^*_\mathrm{GWS}(D) \ge \frac{m-1}{2m}\left(\ln\frac{m-1}{mD}\right)\left(\log\frac{1}{4mD}\right).
\]
These two bounds differ by a factor of $\log\log m$ for distortion $D=\Omega(1/m\log m)$ and a constant factor elsewhere. It is likely that the factor of $\log \log m$ is the result of looseness in our lower bound. On the other hand, the lower bound of Theorem~\ref{thm-dist-lower} gives 
\[
R^*_\mathrm{WS}(D) \ge \frac{1}{2}\left(\log\frac{1}{\sqrt{D}+1/m}\right)\left(\log\frac{1}{4mD}\right),
\]
which is also a lower bound on $R^*_\mathrm{GWS}(D)$. The above two lower bounds only differ by a constant factor for $D=\Omega(m^{-c})$ and $c>0$ and by a factor of $\log(1/D)/\log m$ for $D=o(m^{-c})$ and $c>0$. 

\item For the star network considered in Subsection~\ref{sec-star}, $\lambda_2= 1-1/2(m-1)$ and the upper bound gives
\[
R^*_\mathrm{GWS}(D) \le \frac{2(m-1)}{m}\left(\ln\frac{2}{D}\right)\left(\log\frac{2(m-1)\ln(2/D)}{m^2D}\right).
\]
On the other hand, the upper bound in Subsection~\ref{sec-star} gives
\[
R^*(D) \le \frac{m-1}{m}\log\left(\frac{2(m-1)^2}{m^3D}\right).
\]
These two bounds differ by a factor of $(\log\log m) \log m$ for distortion $D=\Omega(1/m\log m)$ and $\log m$ for $D=o(1/m\log m)$, $D=\Omega(m^{-c})$, and $c>0$. The $\log m$ factor represents the penalty of using the gossip-based distributed protocols.
\end{enumerate}

\subsection{Network Rate-Distortion Function with Side Information}
\label{sec-gossip-wz}

\begin{figure}[!h]
\centering
\small
\psfrag{labeld}[t]{$D$}
\psfrag{labelr}[b]{$R$ bits}
\psfrag{legendsimulation}{Simulation}
\psfrag{legendsimulationwithwynerzivcoding}{Simulation (Wyner-Ziv)}
\psfrag{legendupperbound}{Upper bound (Theorem~\ref{thm-dist-upper})}
\includegraphics[width=5in]{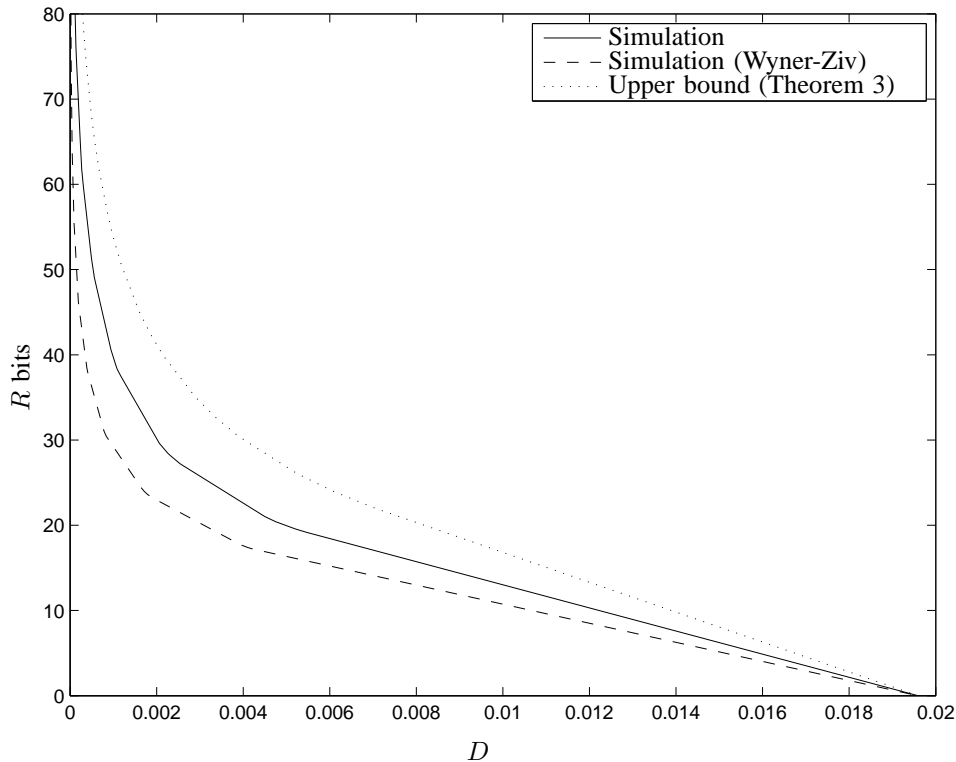}
\caption{Comparison of estimated expected network rate-distortion function $R^*_\mathrm{GWS}(D)$ with and without Wyner-Ziv coding to the upper bound in Theorem~\ref{thm-dist-upper} for $m=50$.}
\label{fig-wz}
\end{figure}

In Subsections~\ref{sec-dist-lower} and~\ref{sec-gossip-upper}, we did not consider the correlation (side information) between the node estimates in computing the transmission rate. While Theorem~\ref{thm-dist-upper} remains an upper bound when side information is considered, the lower bounds in Theorems~\ref{thm-dist-lower} and~\ref{thm-dist-upper} do not necessarily hold. To explore the effect of side information on rate, we consider the gossip-based weighted-sum protocol $(T,Q,R,n,d)$ with Wyner-Ziv coding. Suppose that edge $\{i,j\}$ is selected in round $t+1$. By the Wyner-Ziv theorem, the transmission rate in each round can be reduced from $(1/2)\log (1/d)$ to
\[
r = \frac{1}{2}\log\left(\frac{1-(1-d)\left(\E(S_i(t)S_j(t))\right)^2/\E\left(S_i(t)^2\right)\E\left(S_j(t)^2\right)}{d}\right),
\]
Figure~\ref{fig-wz} compares the simulated $R^*_\mathrm{GWS}(D)$ with and without Wyner-Ziv coding to the upper bound in Theorem~\ref{thm-dist-upper} for a complete graph with $m=50$. Note that the improvement in rate drops from $35\%$ at distortion $D\le 0.0004$ to around $15\%$ for high distortion $0.004\le D <0.02$.

\section{Conclusion}
We introduced a lossy source coding formulation for the distributed averaging problem. We established $R^*(D)$ for the 2-node network with correlated WGN sources and a general cutset lower bound for independent WGN sources. The cutset bound is achieved within a factor of 2 using a centralized protocol in a star network. We then established a lower bound on the network rate distortion function for the class of distributed weighted-sum protocols and bounds on the network rate distortion function for gossip-based weighted-sum protocols. The bounds differ by a factor of only $\log\log m$ for a complete graph network. The results provide insights into the fundamental limits on distributed averaging and on the penalty of using a distributed protocol. 

There many questions that would be interesting to explore. For example: (i) We showed that the cutset bound can be improved for tree networks. Can it be improved in general, or even for simple networks with loops such as a ring? 
(ii) Is the $\log\log m$ factor in the upper bound for the gossip-based weighted-sum protocols necessary? Can the lower bound be tightened? 
(iii) We have investigated distributed weighted-sum protocols with a time-invariant normalized local distortion $d$. Can the order of the rate be reduced by letting $d$ vary with time? (iv) The distributed weighted-sum protocols as defined in the paper do not take advantage of the build-up of correlation in the network. Using Wyner-Ziv coding can indeed reduce the rate as demonstrated in Subsection~\ref{sec-gossip-wz}. It would be interesting to find bounds with side information.

\section{Acknowledgments}
The authors would like to thank Andrea Montanari for pointing out the improved cutset bound on $R^*(D)$ for tree networks. We are also indebted to the anonymous reviewers and to Ioannis Kontoyiannis for valuable comments and suggestions that have greatly helped for improving the presentation of the results in this paper.

\appendix[Proof of Lemmas and Propositions]
\begin{IEEEproof}[Proof of the converse of Propersition~\ref{lm-2node-lower}]
In each time slot, the index sent by node 1 is a function of the source $X_{11}^n$ and the indices sent by node 2 in past time slots. Let the $W_1$ and $W_2$ be the collections of indices sent by nodes 1 and 2 over all time slots, respectively, and consider
\begin{align}
nR_1 &\ge H(W_1) \ge H(W_1|X_{21}^n) \nonumber \\
&= I(X_{11}^n;W_1|X_{21}^n) \nonumber \\
&= \sum_{k=1}^n \left(h(X_{1k}|X_{11}^{k-1},X_{21}^n)-h(X_{1k}|X_{11}^{k-1},X_{21}^n,W_1)\right) \nonumber \\
&\ge \sum_{k=1}^n \left(h(X_{1k}|X_{2k})-h(X_{1k}|X_{2k},X_2^{k-1},{X_2}_{k+1}^n,W_1)\right) \nonumber \\
\label{equ-2node-r1}
&= \frac{n}{2}\log\left(2\pi e\left(P_1-\frac{\sigma_{12}^2}{P_2}\right)\right) -\sum_{k=1}^nh(X_{1k}|X_{2k},U_{1k}),
\end{align}
where $U_{1k} =(W_1,X_2^{k-1},{X_2}_{k+1}^n)$.

To bound the second term, we consider the distortion between the estimate $g_{21}^n$ of node 2 and the weighted-sum $g^n$. The estimate $g_{21}^n$ is a function of $(W_1,X_{21}^n)=(U_{1k},X_{2k})$, and the distortion is equal to
\begin{align*}
\frac{1}{n}\sum_{k=1}^n\E\left((g(X_{1k},X_{2k}) - g_{2k}(U_{1k},X_{2k}))^2\right) 
&\ge \frac{1}{n}\sum_{k=1}^n\E\left(\Var(g(X_{1k},X_{2k})|U_{1k},X_{2k})\right) \\
&\ge \frac{1}{n}\sum_{k=1}^n\E\left(\Var(a_1X_{1k}+a_2X_{2k}|U_{1k},X_{2k}))\right) \\
&= \frac{1}{n}\sum_{k=1}^na_1^2\E\left(\Var(X_{1k}|U_{1k},X_{2k}))\right). 
\end{align*}
Let the average distortion between $g_{21}^n$ and $g^n$ be $D_2$. We can bound the second term in (\ref{equ-2node-r1}) as 
\begin{align*}
\sum_{k=1}^nh(X_{1k}|U_{1k},X_{2k}) &\le \sum_{k=1}^n \frac{1}{2}\log\left(2\pi e\E\left(\Var(X_{1k}|U_{1k},X_{2k})\right)\right) \\
&\le \frac{n}{2}\log\left(2\pi e\sum_{k=1}^n\frac{1}{n}\E\left(\Var(X_{1k}|U_{1k},X_{2k})\right)\right) \\
&\le \frac{n}{2}\log\left(2\pi e\frac{D_2}{a_1^2}\right).
\end{align*}
The transmission rate of node 1 is lower bounded by
\[
R_1 \ge \frac{1}{2}\log\left(\frac{a_1^2(1-\rho^2)P_1}{D_2}\right).
\]
Similarly, let the average distortion between $g_{11}^n$ and $g^n$ be $D_1$, then
\[
R_2 \ge \frac{1}{2}\log\left(\frac{a_2^2(1-\rho^2)P_2}{D_1}\right).
\]
The per-node transmission rate is
\begin{align*}
R &= \frac{1}{2}\left(R_1+R_2\right) \ge \frac{1}{2}\log\left( \frac{a_1a_2(1-\rho^2)\sqrt{P_1P_2}}{\sqrt{D_1D_2}} \right).
\end{align*}
The network rate distortion function is lower bounded by the per-node transmission rate minimized over all distortions $D_1$ and $D_2$ satisfying $\left(D_1+D_2\right)/2\le D$. The minimum is achieved for $D_1=D_2=D$, and thus the lower bound is
\begin{equation*}
R^*(D) \ge \frac{1}{2}\log\left( \frac{a_1a_2 (1-\rho^2) \sqrt{P_1P_2}}{D} \right).
\end{equation*}
\end{IEEEproof}

\begin{IEEEproof}[Proof of Lemma~\ref{lm-gossip-mat}]
\begin{itemize}[\IEEEsetlabelwidth{(ii)}]
\item[(i)] By the assumption that $\Yv$ is independent of $A(t)$, we have
\begin{align*}
\E\left(\|A(t)\Yv\|^2\right) &= \E\left(\Yv^TA(t)^TA(t)\Yv\right) \\
&= \E\left(\E\left(\left.\Yv^TA(t)^TA(t)\Yv\right|\Yv\right)\right) \\
&= \E\left(\Yv^T\E\left(A(t)^TA(t)\right)\Yv\right) \\
&= \E\left(\Yv^TA\Yv\right) \\
&= \E\left((\Yv-J\Yv+J\Yv)^TA(\Yv-J\Yv+J\Yv)\right) \\
&= \E\left((\Yv-J\Yv)^TA(\Yv-J\Yv)\right) + \E\left(\Yv^TJ^TAJ\Yv\right) \\
&\le \lambda_2(A)\E\left((\Yv-J\Yv)^T(\Yv-J\Yv)\right) + \E\left(\Yv^TJ^TJ\Yv\right) \\
&= \lambda_2(A)\E\left(\|\Yv-J\Yv\|^2\right) + \E\left(\|J\Yv\|^2\right). 
\end{align*}
\item [(ii)] We consider the norm squared of $A(t)\Yv-J\Yv$
\begin{align*}
\E\left(\|A(t)\Yv-J\Yv\|^2\right) &= \E\left((A(t)\Yv-J\Yv)^T(A(t)\Yv-J\Yv)\right) \\
&= \E\left((\Yv-J\Yv)^TA(t)^TA(t)(\Yv-J\Yv)\right) \\
&= \E\left(\|A(t)(\Yv-J\Yv)\|^2\right).
\end{align*}
By (i) and $J(\Yv-J\Yv)=\zerov$, we have
\[
\E\left(\|A(t)(\Yv-J\Yv)\|^2\right) \le \lambda_2(A)\E\left(\|\Yv-J\Yv\|^2\right).
\]
This completes the proof.
\end{itemize}
\end{IEEEproof}

\bibliographystyle{IEEEtran}
\bibliography{refs}

\end{document}